\documentclass{article}

\pdfoutput=1

\usepackage{jheppub}

\usepackage{amsmath}
\usepackage{amssymb}
\usepackage{latexsym}
\usepackage{amsfonts}
\usepackage{graphicx}
\usepackage{epstopdf}
\usepackage{mdwlist}
\usepackage{enumitem}
\usepackage{stackengine}
\usepackage{verbatim}

\usepackage{array,multirow}

\usepackage{color}%
\def\hhref#1{\href{http://arxiv.org/abs/#1}{arXiv:#1}} 

\begin{document}


\title{Resurgence, Painlev\'e Equations and Conformal Blocks}

\author{Gerald V. Dunne}
 \affiliation{Physics Department, University of Connecticut, Storrs CT 06269.}
\emailAdd{gerald.dunne@uconn.edu}

\date{\today}

\abstract
{We discuss some physical consequences of the resurgent structure of Painlev\'e equations and their related conformal block expansions.
The resurgent structure of Painlev\'e equations is particularly transparent when expressed in terms of physical  conformal block expansions of the associated tau functions. Resurgence produces an intricate network of inter-relations; some between expansions around different critical points, others between expansions around different instanton sectors of the expansions about the same critical point, and others between different non-perturbative sectors of associated spectral problems, via the Bethe-gauge and Painlev\'e-gauge correspondences. Resurgence relations exist both for convergent and divergent expansions, and can be interpreted in terms of the physics of phase transitions. 
These general features are illustrated with three physical examples: correlators of the 2d Ising model,  the partition function of the Gross-Witten-Wadia matrix model, and the full counting statistics of one dimensional fermions, associated with Painlev\'e VI, Painlev\'e III and Painlev\'e V, respectively.}


\maketitle


\section{Introduction}
\label{sec:intro} 

The Painlev\'e equations are universal nonlinear special functions, appearing in a wide variety of problems in physics, playing an analogous role in nonlinear systems to the familiar special functions of linear problems \cite{ince,mccoy-wu,gromak,clarkson,fokas,mason,forrester-book,Chakravarty:1990gi,Zamolodchikov:1994uw,tw}. Resurgence is a universal feature of the asymptotics of natural mathematical problems, those arising from an underlying physical system described by a set of equations:  differential, integral, difference, algebraic, or functional equations, etc... \cite{Dingle:1973,ecalle,costin-book,sauzin,costin-odes,costin-dmj,Garoufalidis:2010ya,Aniceto:2013fka,Dorigoni:2014hea,gokce}. Recently, all-orders expansions of Painlev\'e tau functions in terms of classical $c=1$ conformal blocks have been found  \cite{Gamayun:2012ma,Gamayun:2013auu,Iorgov:2014vla}.  These Painlev\'e tau function expansions have another physical interpretation in terms of $c=\infty$ conformal blocks \cite{litvinov}, and represent special cases of more general relations for gauge theory partition functions, via blow-up equations \cite{nikita}. Here we argue that these physical conformal block expansions provide a natural way to view the resurgence properties of Painlev\'e equations. The goal is to explore different physical manifestations of the resurgent structure of  Painlev\'e equations.  From the physics perspective, one of the most interesting consequences of resurgence  is that the fluctuations about different sectors of a problem may be related to one another in quantitative ways. This enables predictions about perturbative physics from non-perturbative physical information, and vice versa,  about connections between strong and weak coupling,  about phase transitions from expansions about other parametric regions, and also provides a unified characterization of perturbative/non-perturbative relations in associated quantum spectral problems.

Here we discuss three different (but related) manifestations of resurgence in physical systems described by Painlev\'e equations:
\begin{enumerate}
\item Perturbative fluctuation expansions about {\it different} critical points (in the physical variable) are related to one another in a concrete quantitative way.

\item For local expansions in the vicinity of a {\it given} critical point, perturbative fluctuations about different non-perturbative sectors of the trans-series expansion are related to one another in a concrete quantitative way.

\item
Local expansions about {\it moveable singularities} (those associated with boundary conditions) imply explicit  resurgent relations between different instanton sectors of associated spectral problems.

\end{enumerate}
 
We begin with the Painlev\'e VI equation, in part because of its inherent physical interest (for example, for the Ising model and conformal block expansions), but also because  the other lower Painlev\'e equations are obtained from Painlev\'e VI by known cascades of coalescence of singularities. At one level this implies that  results for Painlev\'e VI flow down to all the other Painlev\'e equations, but interesting new features arise in the lower Painlev\'e systems because the coalescence of singularites can change the character of the singularities (for example, from regular to irregular). This results in a richer structure of resurgent relations, and has significant physical consequences. For example, while the expansions about the fixed regular singularities of the Painlev\'e VI equation, at $t=0$, $t=1$ and $t=\infty$,  are all convergent (and yet have  full trans-series expansions), for the Painlev\'e III and V equations the expansion about the regular singularity at $t=0$ is convergent, while the expansion about the irregular singularity at $t=\infty$ is divergent. Physically, a trans-series may transmute from one form to another across a phase transition. We illustrate this phenomenon with the physical example of the Gross-Witten-Wadia unitary matrix model, the physical properties of which are described by the Painlev\'e III  equation, and in the double-scaling limit by Painlev\'e II. We also discuss the physical meaning of resurgence for the full-counting statistics of one-dimensional fermions, a problem related to both Painlev\'e VI and V.

\section{Resurgence in the Painlev\'e VI equation}
\label{sec:p6}

\subsection{Resurgence relating different PVI critical points: Jimbo's tau function expansion}
\label{sec:jimbo}

The Painlev\'e VI (PVI) equation written in standard form \cite{ince,nist-painleve} reads:
\begin{eqnarray}
\partial_t^2 y &=&\frac{1}{2}\left(\frac{1}{y}+\frac{1}{y-1}+\frac{1}{y-t}\right) (\partial_t y)^2-
 \left(\frac{1}{t}+\frac{1}{t-1}+\frac{1}{y-t}\right) \partial_t y \nonumber\\
 && +\frac{y(y-1)(y-t)}{t^2(t-1)^2}
 \left(\alpha+\frac{\beta\, t}{y^2}+\frac{\gamma(t-1)}{(y-1)^2}+\frac{\delta\, t(t-1)}{(y-t)^2}\right)
 \label{eq:p6}
 \end{eqnarray}
 The PVI equation has three fixed regular singularities, which by M\"obius transformations can be placed at $t=0, 1, \infty$, in addition to one moveable pole singularity which characterizes the boundary conditions. 
 Because of the underlying symmetry under fractional linear transformations, the three fixed singularities are essentially equivalent, and a local expansion about one of them can be obtained directly from the local expansion about one of the others \cite{jimbo}. This is the simplest form of resurgence in the PVI equation:  expansions about the {\it different} fixed singularities, $t=0, 1, \infty$, are structurally and quantitatively related to one another.

This can be seen most explicitly for the ``tau function'' for the PVI equation (\ref{eq:p6}). 
Jimbo showed that the tau function expansion near the fixed singularity at $t=0$ has the following (convergent) form \cite{jimbo}:
\begin{eqnarray}
\tau(t)&\sim & ({\rm constant})\times  t^{(\sigma^2-\theta_0^2-\theta_t^2)/4}\left[1+\frac{(\theta_0^2-\theta_t^2-\sigma^2)(\theta_\infty^2-\theta_1^2-\sigma^2)}{8\sigma^2} \, t \right.\nonumber\\
&&\left. -s \frac{(\theta_0^2-(\theta_t-\sigma)^2)(\theta_\infty^2-(\theta_1-\sigma)^2)}{16\, \sigma^2\, (1+\sigma)^2} \, t^{1+\sigma}
-\frac{1}{s} \frac{(\theta_0^2-(\theta_t+\sigma)^2)(\theta_\infty^2-(\theta_1+\sigma)^2)}{16\, \sigma^2\, (1-\sigma)^2} \, t^{1-\sigma} \right. \nonumber\\
&& \left. +\sum_{j=2}^\infty \sum_{|k|\leq j} a_{jk} t^{j-k \, \sigma}\right] \qquad\qquad, \quad t\to 0
\label{eq:tau-jimbo}
\end{eqnarray}
The parameters $\vec{\theta}\equiv (\theta_0, \theta_1, \theta_t, \theta_\infty)$ in (\ref{eq:tau-jimbo}) describe the monodromy about the points $(0, 1, t, \infty)$, and are related to the parameters $(\alpha, \beta, \gamma, \delta)$ of the PVI equation (\ref{eq:p6}) as \cite{jimbo}
\begin{eqnarray}
\alpha=\frac{1}{2}(\theta_\infty -1)^2\quad, \quad \beta=-\frac{1}{2}\theta_0^2\quad, \quad 
\gamma=\frac{1}{2}\theta_1^2\quad, \quad  \delta=\frac{1}{2}(1-\theta_t^2)
\label{eq:jimbo-parameters}
\end{eqnarray}
The expansion (\ref{eq:tau-jimbo}) is effectively a double expansion, in powers of $t$ and also in powers of $t^\sigma$. The parameters $\sigma$ and $s$ in (\ref{eq:tau-jimbo}) are the two boundary condition parameters of the tau function. The parameter $s$ is an instanton counting parameter. Thus,  (\ref{eq:tau-jimbo}) is a trans-series expansion (with $s$ being the trans-series parameter), even though the expansion is convergent. The overall constant factor appears because the tau function is  only defined up to an overall constant [it is related to the PVI solution $y(t)$ via $\frac{d}{dt} \ln \tau(t)$], and in physical applications this constant  is fixed from matching data for the specific problem at hand. Examples are given below: the Ising model for PVI (Section \ref{sec:ising}), the Gross-Witten-Wadia matrix model for PIII (Section \ref{sec:gww}), and the fermionic full-counting statistics for PV (Section \ref{sec:fcs}). 

Jimbo studied the PVI monodromy and connection problems \cite{jimbo}, and showed that similar tau function expansions exist for the other fixed singularities: $t\sim 1$ and $t\sim \infty$. Most importantly,  these other expansions are related to the $t\sim 0$ expansion in  (\ref{eq:tau-jimbo}) simply by suitable transformations of parameters. These other expansions have precisely the same form as in (\ref{eq:tau-jimbo}), but with different boundary condition parameters $\sigma$ and $s$, and also an exchange of the monodromy parameters $\vec{\theta}$  as follows \cite{jimbo,forrester-witte-tau6,forrester-witte-p6-rmt,guzzetti,its}:
\begin{eqnarray}
t\sim 1\quad&:& \qquad \theta_0\quad \longleftrightarrow \quad \theta_1\\
t\sim \infty\quad&:& \qquad \theta_0\quad \longleftrightarrow \quad \theta_\infty
\label{eq:exchange}
\end{eqnarray}
This is a simple manifestation of resurgence: the expansions about the three different fixed critical points, $t=0$, $t=1$ and $t=\infty$, are essentially the same, up to transformation of the parameters. In Jimbo's words \cite{jimbo}: 
\begin{quote}
{\it ``In the case of (PVI), the fixed critical points $t=0, 1$ and $\infty$ play equivalent roles. Hence the result  above  [(\ref{eq:tau-jimbo})] makes it possible to derive a connection formula for the $\tau$ function for (PVI).''}
\end{quote}
This fact has interesting physical implications, as discussed below [Section \ref{sec:ising}] for the 2d Ising model, whose diagonal correlation functions are  tau functions for the PVI equation, with a certain choice of parameters and boundary conditions \cite{jimbo-miwa,forrester-witte-ising,forrester-witte-p6}.
An important caveat \cite{jimbo}, which is realized in the PVI application to the Ising model, is that (\ref{eq:tau-jimbo}) describes the form of the expansion for {\it generic} monodromy and trans-series parameters $\vec{\theta}$  and $(\sigma, s)$. For {\it non-generic} parameters, for example when a parameter vanishes, or some combination of parameters is related to an integer, the expansion may also develop powers of logarithms. See Section \ref{sec:ising-diagonal}.

\subsection{Resurgence Relating Different PVI Instanton Sectors:  $c=1$ Conformal Blocks}
\label{sec:block}

The PVI tau function has another form of resurgence, connecting different instanton sectors in the expansion about a {\it given} singular point. It has relatively recently been shown that Jimbo's small $t$ expansion (\ref{eq:tau-jimbo}) may be extended to all orders, in a closed-form involving sums over $c=1$ conformal blocks \cite{Gamayun:2012ma,Gamayun:2013auu,Iorgov:2014vla,nagoya,its,Gavrylenko:2016zlf,Alekseev:2018kcn}. This remarkable all-orders expansion is a conformal block expansion for $c=1$ conformal field theories. This expansion is convergent, in agreement with general arguments for conformal operator product expansions \cite{Luscher:1975js,Mack:1976pa,Pappadopulo:2012jk}. Nevertheless, we point out here that the conformal block expansion is  in mathematical terms a {\it trans-series} expansion,  reflecting its natural physical interpretation as an instanton expansion \cite{Nekrasov:2002qd}. Furthermore, the structure of the conformal block expansion of the PVI tau function exhibits the property of resurgence in a particularly transparent manner: the  leading [zero instanton] term encodes all higher instanton sectors in a completely explicit way. When converted from the PVI tau function $\tau(t)$ to the PVI solution $y(t)$, this resurgent structure is of course still present, but it is less obvious simply because it becomes scrambled by the non-linear transformation between $\tau(t)$ and $y(t)$.

To explain this fact, we quote the main result of  \cite{Gamayun:2012ma,Gamayun:2013auu,Iorgov:2014vla}. Jimbo's tau function expansion (\ref{eq:tau-jimbo})  at $t\sim 0$ extends to all orders as follows:
\begin{eqnarray}
\tau(t)&\sim & ({\rm constant})\times  \sum_{n=-\infty}^\infty s^n \, C(\vec{\tilde{\theta}}, \tilde{\sigma}+n)\,{\mathcal B}(\vec{\tilde{\theta}}, \tilde{\sigma}+n; t)
\label{eq:tau-all-orders}
\end{eqnarray}
Here $\vec{\tilde{\theta}}=(\tilde{\theta}_0, \tilde{\theta}_1, \tilde{\theta}_t, \tilde{\theta}_\infty)$ are the monodromy parameters  in (\ref{eq:jimbo-parameters}), and the pair $\{\tilde{\sigma}, s\}$ are the two boundary condition parameters associated with the expansion about $t\sim 0$.
(There is a  change of notational convention between Jimbo's paper \cite{jimbo} and the papers  \cite{Gamayun:2012ma,Gamayun:2013auu}: 
$\vec{\theta}\to \vec{\tilde{\theta}}=\frac{1}{2}\vec{\theta}$, and $\sigma\to \tilde{\sigma}=\frac{1}{2}\sigma$.)
The coefficients $C(\vec{\tilde{\theta}}, \tilde{\sigma})$ are expressed in terms of the Barnes G-function (double gamma function):
\begin{eqnarray}
C(\vec{\tilde{\theta}}, \tilde{\sigma})= \frac{\prod_{\epsilon, \epsilon^\prime =\pm} 
G(1+\tilde{\theta}_t+\epsilon \tilde{\theta}_0 +\epsilon^\prime \tilde{\sigma})G(1+\tilde{\theta}_1+\epsilon \tilde{\theta}_\infty +\epsilon^\prime \tilde{\sigma})}
{G(1+2 \tilde{\sigma}) G(1-2\tilde{\sigma})}
\label{eq:cn}
\end{eqnarray}
The $t$ dependence of the expansion (\ref{eq:tau-all-orders}) resides in the function ${\mathcal B}(\vec{\tilde{\theta}}, \tilde{\sigma}; t)$, which  is the general $c=1$ conformal block, given in closed form as:
\begin{eqnarray}
{\mathcal B}(\vec{\tilde{\theta}}, \tilde{\sigma}; t)=t^{\tilde{\sigma}^2-\tilde{\theta}_0^2-\tilde{\theta}_t^2}(1-t)^{2\tilde{\theta}_t \tilde{\theta}_1} 
\sum_{\lambda, \mu\in {\mathcal Y}} {\mathcal B}_{\lambda, \mu}(\vec{\tilde{\theta}}, \tilde{\sigma}) \, t^{|\lambda|+|\mu|}
\label{eq:block}
\end{eqnarray}
where the sum is over pairs of Young diagrams, $\lambda$ and $\mu$, of size $|\lambda|$ and $|\mu|$, and with combinatorial coefficients involving the hook lengths, $h_\lambda(i, j)$, of elements of the  Young diagrams:
\begin{eqnarray}
 {\mathcal B}_{\lambda, \mu}(\vec{\tilde{\theta}}, \tilde{\sigma})&=& 
 \prod_{(i, j)\in \lambda} \frac{\left(\left(\tilde{\theta}_t+\tilde{\sigma}+i-j\right)^2-\tilde{\theta}_0^2\right)\left(\left(\tilde{\theta}_1+\tilde{\sigma}+i-j\right)^2-\tilde{\theta}_\infty^2\right)}{h_\lambda^2(i, j)\left(\lambda_j^\prime-i+\mu_i-j+1+2\tilde{\sigma}\right)^2} \nonumber\\
& \times& \prod_{(i, j)\in \mu} \frac{\left(\left(\tilde{\theta}_t-\tilde{\sigma}+i-j\right)^2-\tilde{\theta}_0^2\right)\left(\left(\tilde{\theta}_1-\tilde{\sigma}+i-j\right)^2-\tilde{\theta}_\infty^2\right)}{h_\mu^2(i, j)\left(\mu_j^\prime-i+\lambda_i-j+1-2\tilde{\sigma}\right)^2} 
\label{eq:young}
\end{eqnarray}
The primes refer to elements of the transposed diagram.

The conformal block expansion in (\ref{eq:tau-all-orders}) is in fact an instanton sum, over both instanton and anti-instanton sectors, with the boundary condition trans-series parameter $s$ being an instanton counting parameter.
The sum  is expressed in terms of Nekrasov partition functions, as the $C(\vec{\tilde{\theta}}, \tilde{\sigma})$ and ${\mathcal B}(\vec{\tilde{\theta}}, \tilde{\sigma}; t)$ factors in the expansion (\ref{eq:tau-all-orders}) are the perturbative and instanton partition functions, respectively, of an  $SU(2)$ $N_f=4$ gauge theory \cite{Nekrasov:2002qd}. Thus the tau function expansion (\ref{eq:tau-all-orders}) has the form of the dual partition functions of Nekrasov and Okounkov \cite{Nekrasov:2003rj,Nekrasov:2002qd,Mironov:2017lgl,Nekrasov:2018pqq}. 

The resurgent interpretation of the remarkable result (\ref{eq:tau-all-orders}) follows immediately. The contributions from different instanton (and anti-instanton) sectors, labeled by the integer $n$, are obtained from that of the zero-instanton sector simply by shifting the boundary condition parameter $\tilde{\sigma}$ as $\tilde{\sigma}\to \tilde{\sigma}+n$. Three things change:
\begin{enumerate}
\item
The  factor $t^{\tilde{\sigma}^2}$ in (\ref{eq:block}) becomes $t^{(\tilde{\sigma}+n)^2}$, and is associated with the trans-series/instanton-counting  factor $s^n$. Note that while instantons  are usually identified with exponential factors such as $e^{-1/t}$, the power-like instanton factors, $t^{\tilde{\sigma}^2}$, are characteristic of the physical effect of {\it complex} instantons \cite{dykhne,Basar:2015xna}. Also note that when mapped back from the tau function $\tau(t)$  to the PVI solution $y(t)$, the dependence on the instanton label $n$ in the exponent becomes linear rather than quadratic.
\item
The multiplicative factor $C(\vec{\tilde{\theta}}, \tilde{\sigma})$ in (\ref{eq:cn}) changes to $C(\vec{\tilde{\theta}}, \tilde{\sigma}+n)$ in the $n$-instanton sector.
\item
The fluctuation coefficients ${\mathcal B}_{\lambda, \mu}(\vec{\tilde{\theta}}, \tilde{\sigma})$ in (\ref{eq:block}), given by (\ref{eq:young}), change to  ${\mathcal B}_{\lambda, \mu}(\vec{\tilde{\theta}}, \tilde{\sigma}+n)$ in the $n$-instanton sector.

\end{enumerate}
We emphasize that the functions $C(\vec{\tilde{\theta}}, \tilde{\sigma})$ and ${\mathcal B}_{\lambda, \mu}(\vec{\tilde{\theta}}, \tilde{\sigma}; t)$ retain the same functional form: one simply shifts the boundary condition parameter $\tilde{\sigma}\to \tilde{\sigma}+n$.
Therefore, it is sufficient to know just the $n=0$ conformal block in (\ref{eq:block}), together with the normalization coefficient in (\ref{eq:cn}), in order to specify the entire trans-series, including the sum over all instanton sectors, and all fluctuations about each instanton sector. In other words, the conformal block expansion (\ref{eq:tau-all-orders}) of the PVI tau function implies that the $n=0$ sector completely determines the entire trans-series, for the full PVI tau function expansion (\ref{eq:tau-all-orders}).

This is the second manifestation of resurgence mentioned in the Introduction: it connects {\it different instanton sectors} in the expansion about a {\it given} fixed singular point, here $t=0$. This can also be combined with the first manifestation, whereby the tau function expansions about the {\it other} fixed singular points, $t=1$ and $t=\infty$, are related to the $t\sim 0$ expansion by transformation of the parameters. So, we see two quite different levels of resurgent structure in the PVI equations. A third level of resurgent structure for PVI is discussed in the next subsection.

\subsection{Resurgent Perturbative/Non-perturbative Relations from PVI: Classical Conformal Blocks}
\label{sec:p6-classical-blocks}

Another manifestation of resurgence in the PVI equation can be seen in recent results concerning certain PVI solutions, classical conformal blocks and an associated Heun equation spectral problem  \cite{litvinov,Lencses:2017dgf}.
  Conformal blocks enter the story again, but in a different way. 
Here we show how this connection between PVI, classical conformal blocks and Heun spectral problems can be understood in terms of resurgent perturbative/non-perturbative relations that have been found in a wide variety of quantum spectral problems 
  \cite{hoe,alvarez-cubic,alvarez-howls-silverstone,alvarez,Dunne:2013ada,Dunne:2014bca,Dunne:2016qix,Misumi:2015dua,Basar:2017hpr,Ito:2018eon}.

\subsubsection{PVI Connection problem and the Heun Equation}
\label{sec:p6-heun}

Consider a solution to the PVI equation (\ref{eq:p6}) with the following boundary conditions specified at one of the fixed singularities, $t=0$, and at the moveable singularity, which we write as $t=x$. The boundary conditions are:
\begin{eqnarray}
y(t)&\sim & -\kappa\, t^\nu+\dots \qquad\qquad\qquad\qquad\qquad ,\qquad  t\to 0
\label{eq:p6-heun-1}\\
y(t)&\sim & \frac{x(1-x)}{(\lambda_4-1)(t-x)} +y_0+\dots \qquad\qquad,\qquad   t\to x
\label{eq:p6-heun-2}
\end{eqnarray}
This solution is specified by two parameters $(\nu, \kappa)$ at the fixed singularity $t=0$, and two parameters $(x, y_0)$ at the moveable singularity at $t=x$. The first of this second set of parameters is the {\it location} $x$ of the moveable singularity, which for PVI is a simple pole, for which  the residue is fixed by the PVI equation. The other parameter, $y_0$, is the next constant that appears in the expansion of $y(t)$ about the pole. This expansion structure is dictated by the PVI equation. The solution is fully determined by connecting the two sets of boundary conditions in (\ref{eq:p6-heun-1}) and (\ref{eq:p6-heun-2}).

This connection problem has a beautiful solution \cite{litvinov} in terms of the parameter pair $(\nu, x)$, which therefore determines $\kappa=\kappa(\nu, x)$ and $y_0=y_0(\nu, x)$ as functions of $(\nu, x)$.
In the approach of \cite{litvinov} it is convenient and natural to define  another notation for the PVI parameters, related to those of (\ref{eq:p6}) as follows:
\begin{eqnarray}
\alpha\leftrightarrow 2\left(\frac{1}{4}-\delta_5\right) \quad; \quad
\beta\leftrightarrow -2\left(\frac{1}{4}-\delta_1\right) \quad&;& \quad
\gamma\leftrightarrow 2\left(\frac{1}{4}-\delta_3\right) \quad; \quad
\delta\leftrightarrow 2\,\delta_2 \nonumber \\
\delta_i\equiv \frac{1}{4}\left(1-\lambda_i^2\right)\quad, \quad i=1, 2, \dots, 5 &&\qquad; \qquad \lambda_5\equiv \lambda_4-1
\label{eq:p6-conversion}
\end{eqnarray}
We first quote the result, and then explain its resurgent consequences and the physical context from which it arises.

The solution to the connection problem makes use of the well-known fact that PVI defines an isomonodromic Hamiltonian system, from which one can construct an associated Lagrangian and action \cite{gromak,fokas,Nekrasov:2011bc}. This action (suitably regularized \cite{litvinov}) plays a key role in the connection problem, and is related to the Yang-Yang functional of integrable models \cite{Nekrasov:2011bc,Lukyanov:2010rn,Lukyanov:2011wd,litvinov}. Define the function $f(\nu, x)$ as the regularized action evaluated on the PVI solution $y(t)$ satisfying the boundary conditions (\ref{eq:p6-heun-1}) and (\ref{eq:p6-heun-2}):
\begin{eqnarray}
f(\nu, x):&=&\left(\delta_\nu-\delta_1-\delta_2\right) \ln x+\left(\delta_1-\delta_2-\delta_3+\delta_4\right) \ln(1-x) +\frac{\nu}{2}\left(\ln\left(\frac{\kappa}{\kappa_0}\right)+\nu\, \ln x\right)\nonumber\\
&& +\int_0^x dt\left(L\left(y, \dot{y}, t\right) -\frac{\nu^2}{4t}-\frac{\lambda_4-1}{2(x-t)}\right)
\label{eq:p6f}
\end{eqnarray}
where $\delta_\nu\equiv \frac{1}{4}\left(1-\nu^2\right)$, and  $\kappa_0=\kappa_0(\nu)$ is defined as
\begin{eqnarray}
\kappa_0(\nu):=\lim_{x\to 0}\left(x^\nu\, \kappa(\nu, x)\right) =\frac{4\,\nu^2}{(\nu-1-\lambda_3+\lambda_4)(\nu-1+\lambda_3+\lambda_4)}
\label{eq:k0}
\end{eqnarray}
Note that this PVI action $f(\nu, x)$ in (\ref{eq:p6f}) is a function of the boundary condition parameter pair $(\nu, x)$, and of the functions $\kappa(\nu, x)$ and $y_0(\nu, x)$.

In \cite{litvinov} it is shown that the PVI action $f(\nu, x)$ satisfies the following differential relations:
\begin{eqnarray}
\frac{\partial f(\nu, x)}{\partial x}&=& \frac{(\lambda_4-1)^2}{x(1-x)}\left(x-y_0(\nu, x)\right)+\frac{(\delta_3-\delta_1+\delta_2-\delta_4)}{x}
+\frac{(\delta_2+\delta_3+\delta_4-\delta_1)}{1-x}
\label{eq:p6-dfdx}
\\
\frac{\partial f(\nu, x)}{\partial \nu}&=& \frac{1}{2} \ln \kappa(\nu, x)-\frac{1}{2}\frac{\partial}{\partial \nu}\left(\nu\, \ln \kappa_0(\nu)\right)
\label{eq:p6-dfdnu}
\end{eqnarray}
A simple but important observation here is that the consistency of the relations (\ref{eq:p6-dfdx},\ref{eq:p6-dfdnu}), $\frac{\partial^2 f(\nu, x)}{\partial x\, \partial \nu}=\frac{\partial^2 f(\nu, x)}{\partial \nu\, \partial x}$, implies the following relation between $y_0(\nu, x)$ and $\kappa(\nu, x)$:
\begin{eqnarray}
\frac{\partial y_0(\nu, x)}{\partial \nu} =-\frac{1}{2(\lambda_4-1)^2} \, x(1-x)\frac{\partial}{\partial x} \ln \kappa(\nu, x)
\label{eq:pnp}
\end{eqnarray}
At this stage, this is a statement about a particular connection problem for PVI, with the matched boundary conditions in (\ref{eq:p6-heun-1}) and (\ref{eq:p6-heun-2}), expressed in terms of the boundary condition parameter pair $(\nu, x)$. The relation to resurgence comes from the identification of this PVI problem with an associated spectral problem for the Heun equation \cite{litvinov}. 

Associate this PVI connection problem (\ref{eq:p6-heun-1}, \ref{eq:p6-heun-2}) with the following spectral problem for the Heun equation, with singularities at $z=0, 1, x, \infty$:
\begin{eqnarray}
\psi^{\prime\prime}(z)+\left(\frac{\delta_1}{z^2}+\frac{\delta_2}{(z-x)^2}+\frac{\delta_3}{(z-1)^2}+\frac{x(x-1) C}{z(z-1)(z-x)}-\frac{(\delta_1+\delta_2+\delta_3-\delta_4)}{z(z-1)}\right)\psi(z)=0\nonumber\\
\label{eq:heun-eq}
\end{eqnarray}
The Heun parameters $(\delta_1, \delta_2, \delta_3, \delta_4)$ in (\ref{eq:heun-eq}) are identified with the PVI parameters via the relations in  (\ref{eq:p6-conversion}), and the Heun singularity parameter $x$ is identified with the location of the moveable pole for the PVI solution $y(t)$  in (\ref{eq:p6-heun-2}). The PVI boundary condition parameter $\nu$  in (\ref{eq:p6-heun-1}) is identified with the monodromy parameter $\nu$ of the  Heun solution $\psi(z)$. With the prescribed monodromy,  the Heun ``accessory parameter'' $C=C(\nu, x)$  in (\ref{eq:heun-eq}) becomes a function of $\nu$ and $x$. The remarkable result of \cite{litvinov} is that the Heun accessory parameter $C(\nu, x)$ is directly related to the regularized PVI action $f(\nu, x)$ in (\ref{eq:p6f}) as:
\begin{eqnarray}
C(\nu, x)=\frac{\partial}{\partial x} f(\nu, x)
\label{eq:cf}
\end{eqnarray}

This surprising connection between the  PVI connection problem (\ref{eq:p6-heun-1}, \ref{eq:p6-heun-2}) and the Heun equation (\ref{eq:heun-eq}) arises in the following way in the context of classical conformal blocks \cite{litvinov}. Express the  conformal field theory (CFT) central charge $c$ as $c=1+6\left(b+\frac{1}{b}\right)^2$, and consider the classical $c\to\infty$ limit via a $b\to 0$ limit. Then the 5-point CFT correlator with the  vertex operator insertion $V_{(1,2)}(z)$ behaves as \cite{Belavin:1984vu,Zamolodchikov:1995aa,agt,neitzke,Gaiotto:2009ma,fateev,Alba:2010qc,Fitzpatrick:2016ive}
\begin{eqnarray}
\langle V_{(1, 2)}(z) V_{\Delta_1}(0) V_{\Delta_2}(x) V_{\Delta_3}(1) V_{\Delta_4}(\infty)\rangle_{b\to 0} \sim \psi(z; x)\exp\left[\frac{1}{b^2}\, f(\nu, x)\right]
\label{eq:v12}
\end{eqnarray}
where $\psi(z; x)$ satisfies the Heun equation (\ref{eq:heun-eq}), with the classical operator dimension parameters scaled as: $\delta_i\sim b^2\, \Delta_i$. The exponent $f(\nu, x)$ in (\ref{eq:v12}) is the same PVI regularized action in (\ref{eq:p6f}), and  is related to the Heun accessory parameter $C(\nu, x)$ via (\ref{eq:cf}). The connection with PVI arises because in the classical limit, $b\to 0$, the 5-point CFT correlator with the  other vertex operator insertion $V_{(2,1)}(z)$ behaves instead as
\begin{eqnarray}
\langle V_{(2, 1)}(y) V_{\Delta_1}(0) V_{\Delta_2}(t) V_{\Delta_3}(1) V_{\Delta_4}(\infty)\rangle_{b\to 0}  \sim \exp\left[\frac{1}{b^2}\, S(y, t)\right]
\label{eq:v21}
\end{eqnarray}
The exponent $S(y, t)$, with a suitable shift \cite{litvinov}, satisfies a Hamilton-Jacobi equation whose Hamiltonian is identified with that of PVI. Thus, the corresponding classical equation of motion is precisely the PVI equation, with the identifications given above. Another complementary way to understand the relation between PVI and the Heun equation is to recall that each has a representation in terms of Weierstrass functions, in which form we recognize the PVI equation as the classical equation of motion corresponding to the 
Schr\"odinger-like Heun equation. See the Appendix: Section \ref{app:p6-heun-appendix}.

\subsubsection{Interpretation of (\ref{eq:cf}) as a Resurgent  Perturbative/Non-Perturbative Relation}
\label{sec:p6-pnp}

Using the differential relation (\ref{eq:p6-dfdx}), we can convert the remarkable correspondence (\ref{eq:cf}) into a linear relation between the Heun accessory parameter $C(\nu, x)$ and the PVI boundary condition parameter $y_0(\nu, x)$ in the vicinity of the moveable pole $x$, as in (\ref{eq:p6-heun-2}):
\begin{eqnarray}
y_0(\nu, x)= -\frac{x(1-x)}{(\lambda_4-1)^2}\, C(\nu, x) +\frac{x}{2}\left(\frac{\lambda_4-3}{\lambda_4-1}\right)+\frac{(\lambda_1^2-\lambda_2^2-\lambda_3^2+\lambda_4^2)}{4(\lambda_4-1)^2}
\label{eq:y0c}
\end{eqnarray}
In the Heun problem, $C(\nu, x)$ is a spectral eigenvalue, determined by the monodromy parameter $\nu$ and the pole location $x$. 
When translated into Schr\"odinger like form, the pole location $x$ is identified with (the inverse of) a semiclassical parameter, and the monodromy parameter  $\nu$ is identified with the label of the spectral eigenvalue.

This generalizes to PVI an earlier result for a particular Painlev\'e III (PIII) system, the radial Sinh-Gordon equation, which is identified in this way with the spectral problem for the Mathieu equation \cite{novokshenov,Lukyanov:2010rn,Lukyanov:2011wd}. Consider the PIII equation
\begin{eqnarray}
\ddot{y}+\frac{1}{t}\dot{y}=\frac{(\dot{y})^2}{y}+\frac{1}{t}\left(\alpha \, y^2+\beta\right)+\gamma \, y^3+\frac{\delta}{y}
\label{eq:p3-general}
\end{eqnarray}
with the parameter choice $\alpha=\beta=0$, and $\gamma=-\delta=\frac{1}{4}$. In terms of $y(t)\equiv e^{u(t)}$, this becomes the 2 dimensional radial Sinh-Gordon equation:
\begin{eqnarray}
\ddot{u}+\frac{1}{t}\,\dot{u}=\frac{1}{2}\,\sinh(2u)
\label{eq:p3-shg}
\end{eqnarray}
In \cite{novokshenov,Lukyanov:2011wd} it is shown that the following PIII connection problem is directly related to the Mathieu equation spectral problem  (a special reduction of Heun). 
The PIII equation (\ref{eq:p3-shg}) has a regular singular point at $t=0$. We choose boundary conditions at $t=0$ and at a moveable pole location $t=r$:
\begin{eqnarray}
y(t)&\sim & \kappa\, t^{2\nu-1}+\dots \qquad\qquad\qquad\qquad\qquad\qquad\qquad ,\qquad  t\to 0
\label{eq:p3m-1}\\
y(t)&\sim & \frac{2}{(t-r)} -\frac{1}{r}+\frac{(10-16c)}{12r^2} (t-r)+\dots \qquad\qquad,\qquad   t\to r
\label{eq:p3m-2}
\end{eqnarray}
There is one pair of boundary condition parameters, $(\nu, \kappa)$, at $t\sim 0$, and another pair $(r, c)$ at the moveable pole $t\sim r$. These boundary conditions are the PIII analogue of the PVI boundary conditions in (\ref{eq:p6-heun-1}, \ref{eq:p6-heun-2}). [We use the notation of \cite{Lukyanov:2011wd}, but note that there equations (4.30) and (4.33) are written in terms of the {\it square} of the PIII solution, $(y(t))^2$.] As in the PVI case above, we choose to connect these two sets of boundary conditions in terms of the pair $(\nu, r)$, which therefore determines $\kappa=\kappa(\nu, r)$ and $c=c(\nu, r)$. In \cite{Lukyanov:2011wd} it is shown that the resulting on-shell PIII action $\mathcal S^*(\nu, r)$ satisfies the differential relations:
\begin{eqnarray}
r\frac{\partial \mathcal S^*(\nu, r)}{\partial r}&=&-c(\nu, r)-\frac{r^2}{8}+\frac{1}{4}
\label{eq:p3-dsdr}\\
\frac{\partial \mathcal S^*(\nu, r)}{\partial \nu}&=&\ln \kappa(\nu, r)
\label{eq:p3-dsdnu}
\end{eqnarray}
These should be compared with the differential relations (\ref{eq:p6-dfdx},\ref{eq:p6-dfdnu}) in the PVI connection problem, with the identifications: $x\leftrightarrow r$, $c\leftrightarrow y_0$, and  $\mathcal S^*\leftrightarrow f$. The consistency of (\ref{eq:p3-dsdr}, \ref{eq:p3-dsdnu}) implies the relation [equation (4.34) in  \cite{Lukyanov:2011wd}]:
\begin{eqnarray}
\frac{\partial c(\nu, r)}{\partial \nu} = -r\, \frac{\partial \ln \kappa(\nu, r)}{\partial r}
\label{eq:p3-magic}
\end{eqnarray}
Then in the small $r$ limit one finds the following expansion
\begin{eqnarray}
c(\nu, r)=\nu^2+\frac{1}{2(\nu^2-1)}\left(\frac{r}{4}\right)^4
+\frac{5\nu^2+7}{32(\nu^2-1)^3(\nu^2-4)}\left(\frac{r}{4}\right)^8+\dots
\label{eq:mathieu-c}
\end{eqnarray}
Lukyanov points out that this is the formal large-$\hbar$ expansion for the center of the $N^{\rm th}$ gap in the spectrum of the Schr\"odinger-like Mathieu equation
\begin{eqnarray}
-\frac{\hbar^2}{2} \frac{d^2 \psi}{d z^2}+\cos(z)\, \psi=E\, \psi
\label{eq:mathieu}
\end{eqnarray}
with the identifications:
\begin{eqnarray}
r=\frac{8}{\hbar} \qquad, \qquad E=\frac{\hbar^2}{8}\, c\qquad, \qquad \nu=N
\label{eq:mathieu-id}
\end{eqnarray}
Thus the moveable pole location $r$ is identified with the inverse semiclassical parameter, and the 
boundary condition parameter $c(\nu, r)$ at the moveable pole location is identified with the energy eigenvalue, with the normalizations listed above. The monodromy parameter $\nu$ is identified with the label $N$ of the spectral gap. 

The spectral interpretation of the remaining boundary condition parameter, $\kappa(\nu, r)$, was not identified in \cite{novokshenov,Lukyanov:2011wd}, but by comparison with the full non-perturbative energy spectrum, including the all-orders gap splitting, we see that $\kappa(\nu, r)$ is directly related to the non-perturbative width of the $N^{\rm th}$ gap, with $\nu=N$ \cite{Basar:2015xna,Dunne:2016qix,Gorsky:2017ndg}. Indeed, with the above identifications, we can re-express (\ref{eq:p3-magic}) as
\begin{eqnarray}
\kappa(\nu, \hbar)=\tilde{\kappa}(\nu)\, \exp\left[8\int^\hbar \frac{d\hbar}{\hbar^3} \frac{\partial E(\nu, \hbar)}{\partial \nu}\right]
\label{eq:kappa}
\end{eqnarray}
where $\tilde{\kappa}(\nu)$ is an $\hbar$-independent normalization factor. We recognize this as the perturbative/non-perturbative relation for the Mathieu system, which relates the non-perturbative gap splitting $\kappa(\nu, \hbar)$ to the perturbative gap center $E(\nu, \hbar)$, for the Mathieu spectral problem (\ref{eq:mathieu}), in the large $\hbar$ gap regime. This relation can also be expanded in the small $\hbar$ regime, yielding the {\it band} locations and non-perturbative widths, with the identification $\nu=\left(N+\frac{1}{2}\right)$ for the $N^{\rm th}$ spectral band.
Thus, the full spectral properties of the Mathieu problem are encoded in the connection problem for the PIII equation (\ref{eq:p3-shg}) with the boundary conditions (\ref{eq:p3m-1},\ref{eq:p3m-2}). The relation (\ref{eq:p3-magic}) is the perturbative/non-perturbative (P/NP) relation of the Mathieu system, which connects the perturbative and non-perturbative spectral data  \cite{Basar:2015xna,Dunne:2016qix}.
\begin{table}[htb]
\centering
 \medskip
\begin{tabular}{| c | c | }\hline
Painlev\'e III  boundary condition parameters & Mathieu spectral parameters \\ \hline 
  moveable pole location: $r$ & inverse semiclassical parameter $\frac{1}{\hbar}$ \\\hline
expansion constant at $r$: $c$   & perturbative eigenvalue  \\ \hline
exponent $\nu$ at  fixed singularity $t=0$  & monodromy parameter (gap or band label) \\ \hline
coefficient $\kappa$ at fixed singularity $t=0$ & non-perturbative gap splitting \\ \hline
\end{tabular}
 \medskip
 \caption{Identification between the Painlev\'e III boundary condition parameters in (\ref{eq:p3m-1}, \ref{eq:p3m-2})  and the parameters appearing the associated Mathieu spectral problem (\ref{eq:mathieu}).}
\label{tab:dictionary-mathieu}
\end{table}

Comparing this with the results of \cite{litvinov} for PVI and the Heun equation, we see that the relation (\ref{eq:pnp}) is the generalization of the PIII and Mathieu equation P/NP relation (\ref{eq:p3-magic}): it relates the derivative with respect to the monodromy of the Heun accessory parameter $\mathcal C(\nu, x)$, which  by (\ref{eq:y0c}) is linearly related to the boundary condition parameter $y_0(\nu, x)$ at the moveable pole location $x$, to the derivative with respect to the moveable pole location (which characterizes the inverse semiclassical parameter) of the non-perturbative contribution to the spectral parameter. (Note that the appearance of $x(1-x)$ factors for PVI is characteristic of tau function expressions for PVI, compared to a factor of $x$  for PIII tau functions.)

These identifications have several interesting consequences:
\begin{enumerate}

\item
Explicit P/NP relations, connecting perturbative and non-perturbative expansions, were first found in the quantum oscillator systems for the cubic oscillator and the double-well potential \cite{alvarez-cubic,alvarez-howls-silverstone,alvarez}, and later extended to periodic systems and SUSY models \cite{Dunne:2013ada,Dunne:2014bca,Dunne:2016qix,Misumi:2015dua}. Indeed, such P/NP relations exist for all genus 1 systems \cite{Mironov:2009uv,KashaniPoor:2012wb,Basar:2017hpr,Gorsky:2017ndg,Grassi:2018spf,Ito:2018eon}. The PVI result (\ref{eq:pnp})  encodes all these results via suitable scaling reductions. Note also that the Hamiltonian form of PVI naturally corresponds to genus 1 systems.
This reflects the physics of the Painlev\'e/gauge correspondence in \cite{Zabrodin:2011fk,Bonelli:2016qwg,Grassi:2018spf}. The PVI parameters are associated with masses of the hypermultiplets in ${\mathcal N}=2$ SUSY gauge theory, and mass decoupling cascades are associated with the standard Painlev\'e coalescence cascades.


\item
In the Mathieu system, the band splitting is due to the effect of real instantons, while the gap splitting is due to complex instantons \cite{dykhne,Basar:2015xna}. A similar interpretation for PVI is consistent with the quantum geometry and exact WKB explanations of the P/NP relations for all genus 1 systems, in terms of all-orders actions and dual actions \cite{Mironov:2009uv,KashaniPoor:2012wb,piatek,Krefl:2013bsa,Ashok:2016yxz,Basar:2017hpr,Codesido:2016dld,Gorsky:2017ndg,Bonelli:2016qwg,Grassi:2018spf,Ito:2018eon}.

\item
A similar, but not identical, relation between spectral problems and Painlev\'e equations arises for the PI equation, whose tritronqu\'ee poles have been associated with a leading order WKB analysis of the cubic QM oscillator \cite{Masoero:2010um,novokshenov,novokshenov2}.

\item
Another identification of Painlev\'e equations with spectral boundary value problems
arises in the study of PT symmetric quantum mechanics \cite{Bender:2015bja}. A certain nonlinear eigenvalue problem for Painlev\'e I, II and IV, with prescribed initial value boundary conditions at the origin, with a fixed number of poles, can be associated with a WKB analysis of PT symmetric QM problems, with potentials $V(z)=z^2 (i z)^{\epsilon}$, with $\epsilon=1, 2, 4$, respectively.

\end{enumerate}

\section{Resurgence in the 2D Ising Model: Painlev\'e VI}
\label{sec:ising}

In this Section we discuss some  physical implications of the different manifestations of resurgence in the Painlev\'e VI equation, for the 2 dimensional lattice Ising model. For clarity of presentation we consider the  2d Ising model on the isotropic square lattice, but the discussion generalizes to the anisotropic case. 

\subsection{Resurgence and Kramers-Wannier Duality}

The simplest form of resurgence in the 2d Ising model is the familiar Kramers-Wannier duality \cite{mccoy-wu}. 
For the 2d Ising model, in the absence of an external magnetic field, there are three special temperatures: $T=0$, the critical temperature $T=T_c$, and $T=\infty$. But because of Kramers-Wannier duality, the theory at low $T$ is directly related to the theory at high $T=\infty$. On the 2d  square lattice, the Ising model with inverse temperature $\beta=1/T$ and isotropic coupling $J$, is equivalent to that with  inverse temperature  $\tilde{\beta}$,  where $\beta$ and $\tilde{\beta}$ are related by:
 \begin{eqnarray}
 \tanh(\beta J)=e^{-2\tilde{\beta}\, J}
 \label{eq:kw}
 \end{eqnarray}
 This duality maps high temperature (small $\beta$)  to low temperature (large $\tilde{\beta}$), and vice versa. The self-dual point is when $\beta=\tilde{\beta}$, or in other words, $\sinh(2\beta_c J)=1$. This yields the critical temperature of the 2d Ising phase transition.
 In terms of the partition function on an $N\times N$ isotropic square lattice, Kramers-Wannier duality implies that \cite{kogut}
\begin{eqnarray}
\frac{Z(\beta)}{\sinh^{N/2}(2\beta J)}=\frac{Z(\tilde{\beta})}{\sinh^{N/2}(2\tilde{\beta} J)}
\label{eq:kwz}
\end{eqnarray}
It is convenient for later applications  to Painlev\'e VI to work  in terms of  the variable $t\equiv {\rm sinh}^4(2\beta J)$, in which case the Kramers-Wannier duality is expressed as $t\leftrightarrow 1/t$, with the critical temperature $T_c$ corresponding to the critical value $t_c=1$ (compare with (\ref{eq:kw})):
\begin{eqnarray}
t&=&\sinh^4(2\beta J)= 
\left(\frac{2 \left[\tanh(\beta J)\right]}{\left(1-\left[\tanh(\beta J)\right]^2 \right)}\right)^4 \\
\frac{1}{t}&=&\frac{1}{\sinh^4(2\beta J)}=\left(\frac{2\,  \left[e^{-2\beta J}\right]}{\left(1-\left[e^{-2\beta J}\right]^2\right)}\right)^4
\end{eqnarray}
Indeed, from the exact integral representation of the free energy \cite{mccoy-wu}  for the infinite square lattice, we find the following high and low temperature expansions
\begin{eqnarray}
-\beta\, \mathcal F(t)+\frac{1}{4}\ln t 
&\sim & \frac{1}{8}\ln \frac{1}{t}-\frac{1}{4\,\sqrt{t}}+\frac{1}{32\, t}-\frac{1}{48\, t^{3/2}}+\frac{9}{1024\, t^2}- \frac{9}{1280\, t^{5/2}}+\cdots
\quad, \quad t\to\infty\nonumber\\
&\sim & \frac{1}{8}\ln t-\frac{\sqrt{t}}{4}+\frac{t}{32}-\frac{t^{3/2}}{48}+\frac{9\,t^2}{1024}-\frac{9\,t^{5/2}}{1280}+\cdots\quad, \quad t\to 0 
\label{eq:free-kw-exp}
\end{eqnarray}
These low and high  temperature expansions  have exactly the same form, with identical coefficients. This Kramers-Wannier duality is an example of the first manifestation of resurgence: the formal expansions about the two different asymptotic sectors at $t=0$ and $t=\infty$ are the same. Each expansion is convergent, and the radius of convergence identifies the critical temperature at $t_c=1$. 

In fact, the expansion about the critical temperature can also be deduced from the expansion about $t=0$, or about $t=\infty$. This is a simple consequence of Darboux's theorem (\cite{fisher,henrici, guttmann}, see the Appendix in Section \ref{app:darboux}). The behavior of the expansion about a certain point (e.g., at $t=0$) is directly related to the expansion about a neighboring singularity (here at $t=1$). For example, from the leading large order growth of the coefficients of the (convergent) expansion about $t=0$, we learn that the singularity at the critical point $t_c=1$ is logarithmic, and from subleading corrections to the large-order growth of the coefficients  about $t=0$ we deduce low order terms of the analytic function  multiplying the logarithmic singularity.  This can be seen particularly explicitly for the internal energy:
\begin{eqnarray}
U(t)= -J \sqrt{1+\frac{1}{\sqrt{t}}}\left[1+\frac{2}{\pi} \left(\sqrt{t}-1\right) \mathbb K(t)\right]
\label{eq:internal}
\end{eqnarray}
This clearly has a radius of convergence equal to $1$, as the elliptic integral function $\mathbb K(t)=\frac{\pi}{2}_2F_1\left(\frac{1}{2}, \frac{1}{2}, 1; t\right)$ has a log branch point at $t=1$, but this can be seen directly from the large-order behavior of the expansion coefficients even without knowing the closed-form hypergeometric expression. The  large order behavior of the expansion coefficients, including subleading terms, can be found numerically:
\begin{eqnarray}
c_n=\frac{\left(\left(\frac{1}{2}\right)_n\right)^2}{(n!)^2} \sim \frac{1}{\pi} \left(\frac{1}{n}-\frac{1/4}{n(n-1)}+\frac{9/32}{n(n-1)(n-2)} +\dots\right)
\label{eq:u-large}
\end{eqnarray}
From Darboux's theorem (\ref{eq:darboux3}, \ref{eq:darboux4}), this leading large order growth identifies the  behavior at $t_c=1$ as {\it logarithmic}, and the sub-leading terms determine the fluctuations about this logarithmic singularity:
\begin{eqnarray}
~_2F_1\left(\frac{1}{2}, \frac{1}{2}, 1; t\right)\sim \left(1-\frac{(1-t)}{4}+\frac{9(1-t)^2}{64}+\dots\right)\ln(1-t) +\text{analytic}\quad, \quad t\to 1^-
\label{eq:internal-inversion}
\end{eqnarray}
This reflects the familiar connection properties of the hypergeometric functions, but the more important message is that information about the behavior in the vicinity of the critical point is also encoded in the (convergent) expansion about $t=0$. In other words, "large-order/low-order" resurgence relations are not only present for asymptotic series, but also for convergent series. Indeed, one way to understand "large-order/low-order" resurgence relations  for asymptotic series \cite{berry-howls} is in terms of Darboux's theorem applied to the convergent Borel transform function.

\subsection{Ising Diagonal Correlators and Painlev\'e VI Resurgence}
\label{sec:ising-diagonal}

A deeper manifestation of resurgence in the Ising model  connects {\it different instanton sectors} of the trans-series representation  of the expansions about a {\it given} temperature. This occurs in the 2d Ising model because the diagonal spin-spin correlators are tau functions for Painlev\'e VI, for a particular choice of Painlev\'e VI parameters, and also for a special choice of boundary conditions \cite{jimbo-miwa,forrester-witte-ising}. To see how this richer resurgent structure operates, consider  the diagonal spin-spin correlation function
\begin{eqnarray}
\mathcal C(t, N) \equiv  \langle \sigma_{[0,0]}\, \sigma_{[N, N]}\rangle(t)
\label{eq:ising-correlator}
\end{eqnarray}
This diagonal correlator depends on two variables: the temperature variable $t\equiv \sinh^4(2\beta J)$, and the diagonal spin-separation lattice distance $N$. This correlator  $\mathcal C(t, N)$ is the tau function for PVI with the following choice of parameters \cite{Gamayun:2013auu} (with Jimbo's 
 notation):
\begin{eqnarray}
(\theta_0, \theta_t, \theta_1, \theta_\infty)=(0, N, -N, 1)
\label{eq:p6-ising}
\end{eqnarray}

This representation of the diagonal correlator as a PVI tau function has two immediate consequences concerning resurgent asymptotics in the Ising model. First, Jimbo's result (Section \ref{sec:jimbo}) that the PVI tau function expansions about the different fixed PVI singularities at $t=0$, $t=1$, $t=\infty$ are equivalent up to transformation of parameters \cite{jimbo} implies that there are quantitative relations between the diagonal correlators at low temperature, high temperature and near the critical temperature. However, the Ising PVI parameters and boundary conditions are {\it non-generic}: the boundary condition parameter $\sigma$ in (\ref{eq:tau-jimbo}) vanishes for the Ising  correlators,
which has the physical implication that the expansion about the critical temperature at $t_c=1$ corresponds to the resonant case where logarithmic terms \cite{jimbo} appear in the expansion of the correlators. 
Second, the all-orders extension (Section \ref{sec:block}) of Jimbo's tau function expansion in terms of $c=1$ conformal blocks implies that there are direct resurgent relations between the different instanton sectors of the trans-series expansion of the correlators, when expanded about a {\it given} temperature: $t=0$, $t=1$, or $t=\infty$. Also in this manifestation of resurgence, the non-generic integer parameter $N$ has interesting consequences, which are both simplifications and complications: there are more compact and explicit expressions for the correlators, but  the resurgent interpretation is somewhat more difficult to recognize.

\subsubsection{Toeplitz determinant representation of Ising correlators}
\label{sec:toeplitz}

A compact explicit form of the diagonal Ising correlators $\mathcal C(t, N)$ is in terms of $N\times N$ Toeplitz determinants \cite{wu-toeplitz,forrester-witte-ising,deift}. Define  high and low temperature variables as:
\begin{eqnarray}
t_>\equiv \sinh^4(2\beta J)
\qquad, \qquad 
t_< \equiv \frac{1}{\sinh^4(2\beta J)}
\label{eq:t}
\end{eqnarray}
Then at high temperature the diagonal correlator has the determinant form:
\begin{eqnarray}
\mathcal C^{(+)}(t_>, N)&=& {\rm det}\left(w^{(+)}_{i-j}(t_>)\right)_{i,j = 1,\dots N}
\label{eq:toeplitz-high}
\end{eqnarray}
where the entries of the $N\times N$ matrix are hypergeometric functions \cite{forrester-witte-ising}:
\begin{eqnarray}
w^{(+)}_j(t_>)  &=&\frac{(-1)^{j-1}}{\pi\,\sqrt{t_>}} 
\frac{\Gamma\left(j-\frac{1}{2}\right) \Gamma\left(\frac{1}{2}\right)}{\Gamma(j)}\, 
~_2F_1\left(-\frac{1}{2}, j-\frac{1}{2}, j; t_>\right)
\qquad, \quad j>0 \\
w^{(+)}_{-j}(t_>) &=& \frac{(-1)^{j}\, (t_>)^{j+1/2}}{\pi} \frac{\Gamma\left(j+\frac{1}{2}\right) \Gamma\left(\frac{3}{2}\right)}{\Gamma(j+2)}\, ~_2F_1\left(\frac{1}{2}, j+\frac{1}{2}, j+2; t_>\right)\qquad, \quad j\geq 0
\label{eq:high-w}
\end{eqnarray}
Analogously, at low temperature:
\begin{eqnarray}
\mathcal C^{(-)}(t_>, N)&=& \det\left(w^{(-)}_{i-j}(t_<)\right)_{i,j = 1,\dots N}
\label{eq:toeplitz-low}
\end{eqnarray}
where the entries of the $N\times N$ matrix are other hypergeometric functions:
\begin{eqnarray}
w^{(-)}_j(t_<) &=&\frac{(-1)^{j-1}}{\pi\,(t_<)^{j}} 
\frac{\Gamma\left(j-\frac{1}{2}\right) \Gamma\left(\frac{3}{2}\right)}{\Gamma(j+1)}\, 
~_2F_1\left(\frac{1}{2}, j-\frac{1}{2}, j+1; t_<\right)
\qquad, \quad j>0 \\
w^{(-)}_{-j}(t_<) &=& \frac{(-1)^{j}}{\pi} \frac{\Gamma\left(j+\frac{1}{2}\right) \Gamma\left(\frac{1}{2}\right)}{\Gamma(j+1)}\, ~_2F_1\left(-\frac{1}{2}, j+\frac{1}{2}, j+1; t_<\right)\qquad, \quad j\geq 0
\label{eq:low-w}
\end{eqnarray}
For example, when $N=1$ (just one step along the lattice diagonal),
\begin{eqnarray}
\mathcal C^{(+)}(t_>, 1)&=& \frac{\sqrt{t_>}}{2} ~_2F_1\left(\frac{1}{2}, \frac{1}{2}, 2; t_>\right) = \frac{2}{\pi}\left( \frac{1}{\sqrt{t_>}} \, \mathbb E(t_>) 
+\frac{(t_>-1)}{\sqrt{t_>}} \mathbb K(t_>)\right)
\\
\mathcal C^{(-)}(t_<, 1)&=&  ~_2F_1\left(-\frac{1}{2}, \frac{1}{2}, 1; t_<\right)=\frac{2}{\pi}\, \mathbb E(t_<)
\label{eq:ising-N1}
\end{eqnarray}
When $N=2$ (two steps along the lattice diagonal), 
\begin{eqnarray}
\mathcal C^{(+)}(t_>, 2)&=& \left(\frac{2}{\pi}\right)^2\left(\frac{(5-t_>)}{3 t_>} \mathbb E(t_>)^2 + \frac{8}{3}\frac{(t_>-1)}{t_>} \mathbb E(t_>)  \mathbb K(t_>) 
+\frac{(t_>-1)^2}{t_>} \mathbb K(t_>)^2\right)
\\
\mathcal C^{(-)}(t_<, 2)&=&\left(\frac{2}{\pi}\right)^2\left(\frac{(5t_<-1)}{3t_<} \mathbb E(t_<)^2 + \frac{2}{3}\frac{(t_<-1)^2}{t_<} \mathbb E(t_<)  \mathbb K(t_<) 
 -\frac{(t_<-1)^2}{3 t_<} \mathbb K(t_<)^2\right)
\label{eq:ising-N2}
\end{eqnarray}
For general $N$, the correlators $\mathcal C^{(\pm)}(t, N)$ can be expressed in terms of {\it polynomials} of degree $N$ in the elliptic function ratio $\mathbb E(t)/\mathbb K(t)$  and in $t$ \cite{Ghosh:1984zd}:
\begin{eqnarray}
\mathcal C^{(+)}(t_>, N) &=& \left(\frac{2\,\mathbb K(t_>)}{\pi\, \sqrt{t_>}}\right)^N \text{polynomial}_N\left(\frac{\mathbb E(t_>)}{\mathbb K(t_>)}, t_>\right) 
\nonumber \\
\mathcal C^{(-)}(t_<, N) &=& \left(\frac{2\,\mathbb K(t_<)}{\pi \, \sqrt{t_<}}\right)^N  \text{polynomial}_N\left(\frac{\mathbb E(t_<)}{\mathbb K(t_<)}, t_<\right) 
\end{eqnarray}
This makes it clear that the radius of convergence of the expansions about $t=0$ is 1, which identifies the critical temperature as $t_c=1$. Furthermore, the expansion about the critical temperature, $t_c=1$, can be deduced from the large order behavior of the expansion coefficients about $t=0$, and this identifies $\left(\ln | t-t_c |\right)^k$ behavior, for $k=0, 1, \dots, N$,  on either side of $t_c$. This also follows from the transformation properties of the hypergeometric functions [or those for $\mathbb E(t)$ and $\mathbb K(t)$] under $t\to 1-t$. And for any $N$, the Kramers-Wannier duality under $t\to1/t$ can also be understood via the associated transformation properties of the hypergeometric functions [or those for $ \mathbb E(t)$ and $\mathbb K(t)$] \cite{mccoy-duality}.

Resurgence of the diagonal correlators follows immediately from the fact that for a given $N$ the correlator is just a sum of products of hypergeometric functions, which are themselves  resurgent.
These $N\times N$ Toeplitz determinant representations also imply that the diagonal correlators $\mathcal C(t, N)$ satisfy an $(N+1)^{\rm th}$ order {\it linear} differential equation with respect to $t$ \cite{Boukraa:2006bt}. This implies that for any fixed $N$, the correlators are resurgent functions of $t$ \cite{costin-odes,costin-book}. The diagonal correlators also satisfy a Toda equation which can be used to generate correlators at higher $N$ \cite{Mangazeev:2010vu}:
\begin{eqnarray}
\frac{\partial }{\partial t}\left(t\frac{\partial}{\partial t}\right) \ln \, \mathcal C(t, N)=\frac{N^2}{(1-t)^2}+\frac{\left(N^2-\frac{1}{4}\right)}{(1-t)^2}\frac{\mathcal C(t, N+1) \mathcal C(t, N-1)}{\mathcal C^2(t, N)}
\label{eq:toda}
\end{eqnarray}

\subsubsection{Conformal Block Expansion of Diagonal Ising Correlators}
\label{sec:ising-block}

The full resurgent structure of the Ising model can be seen even more explicitly in terms of the conformal block expansion of the diagonal correlators $\mathcal C^{(\pm)}(t, N)$. Using the conformal block expansion in (\ref{eq:tau-all-orders}), suitably reduced for the special non-generic choice of PVI parameters and boundary conditions, the correlators can be written as a sum over partitions, or Young diagrams. This can be traced to combinatoric results of Gessel and Borodin (see \cite{borodin}). For example, the low temperature expansion of the diagonal correlator can be written as
\begin{eqnarray}
\mathcal C^{(-)}(t_<, N) =\sum_{\lambda\in \mathbb Y, \lambda_1\leq N} \prod_{(i, j)\in \lambda}\frac{\left((i-j)^2-\frac{1}{4}\right)}{h_\lambda^2(i, j)}\, t_<^{|\lambda|}
\label{eq:ising-young}
\end{eqnarray}
where $\mathbb Y$ is the set of all partitions $\lambda=(\lambda_1\geq \lambda_2\geq \dots \geq \lambda_k >0)$, and the sum is restricted to those partitions with $\lambda_1\leq N$. The power of $t_<$ in (\ref{eq:ising-young}) is given by $|\lambda|\equiv \lambda_1+\lambda_2+\dots +\lambda_k$, the total number of boxes in the corresponding Young diagram. The symbol $h_\lambda(i, j)$ is the hook length of the $(i, j)$ box in the Young diagram. 

For example, when $N=1$ the contributing Young diagrams all have just one row, so  
\begin{eqnarray}
\mathcal C^{(-)}(t_<, N=1) &=& \sum_{k=1}^\infty t^k \left(\prod_{j=1}^k\frac{(1-j)^2-\frac{1}{4}}{(k-j)^2} \right)
= 1- \frac{t_<}{4}-\frac{3\, t_<^2}{64} -\frac{5\, t_<^3}{256}-\dots 
\label{eq:ising-young-N1}
\end{eqnarray}
which coincides with the  (convergent) small $t_<$ expansion of the elliptic function representation of $C^{(-)}(t_<, N=1)$  in (\ref{eq:ising-N1}).
When $N=2$,  the contributing Young diagrams  have at most two rows, so there are two kinds of contributions. A simple computation yields
\begin{eqnarray}
\mathcal C^{(-)}(t_<, N=2) &=&1-\frac{t_<}{4} - \frac{3\, t_<^2}{32}-\frac{9\, t_<^3}{256}-\dots 
\label{eq:ising-young-N2}
\end{eqnarray}
which coincides with the (convergent) small $t_<$ expansion of the elliptic function representation of $C^{(-)}(t_<, N=2)$ in (\ref{eq:ising-N2}). A similar expansion can be made for any $N$. Thus, the conformal block expansion (\ref{eq:ising-young}) gives a closed-form combinatorial expression for all orders of the low temperature of the diagonal correlator $\mathcal C^{(-)}(t_<, N)$. A similar argument can be applied to the high temperature expansion.
We comment that the diagonal correlators can alternatively be expressed in terms of generalized hypergeometric functions, which are  written as sums over partitions \cite{forrester-witte-p6-rmt}.

\subsubsection{Generalized Form Factor Expansions as Trans-Series Expansions}
\label{sec:ising-ff}

Another commonly studied expansion for the Ising diagonal correlators is a generalized form factor expansion \cite{jimbo-miwa,Boukraa:2006bt,forrester-witte-ising,trinh}. The diagonal correlators can be expanded as:
\begin{eqnarray}
\text{high temperature :}\quad \mathcal C^{(+)}(t_>, N)&=& \left(1-t_>\right)^{1/4}\sum_{p=0}^\infty f^{(2p+1)}(t_>, N) \\
\text{low temperature :}\quad \mathcal C^{(-)}(t_<, N)&=& \left(1-t_<\right)^{1/4}\left(1+\sum_{p=0}^\infty f^{(2p)}(t_<, N) \right)
\label{eq:ising-ff}
\end{eqnarray}
The lowest two coefficient  functions are:
\begin{eqnarray}
f^{(1)}(t, N)&=&\frac{\left(\frac{1}{2}\right)_{N}}{N!} t^{N/2}  \, _2F_1\left(\frac{1}{2}, N+\frac{1}{2}; N+1; t\right)
\label{eq:ising-f1} \\
f^{(2)}(t, N)&=& t^{N+1}\sum_{j=0}^\infty (j+1) \left(\frac{1}{2}\right)_{j+N} \left(\frac{3}{2}\right)_{j+N}  t^{j} \times \nonumber\\
&&\times \frac{
   _2F_1\left(-\frac{1}{2},j+N+\frac{1}{2};j+N+2;t\right) \,
   _2F_1\left(\frac{1}{2},j+N+\frac{3}{2};j+N+2;t\right)}{4 ((j+N+1)!)^2}
\label{eq:ising-f2}
\end{eqnarray}
There are recursion formulas for generating the higher-order coefficient functions $f^{(k)}(t, N)$.

These expansions can be extended, along the lines of Wu's expansions of horizontal and vertical correlators \cite{wu-toeplitz}, and the Jimbo/Miwa diagonal correlator expansions \cite{jimbo-miwa}, as {\it generalized form factor expansions}, depending on an extra parameter which is commonly written as $\lambda$:
\begin{eqnarray}
\text{high temperature :}\quad \mathcal C^{(+)}(t_>, N; \lambda)&=& \left(1-t_>\right)^{1/4}\sum_{p=0}^\infty \lambda^{2p+1}\, f^{(2p+1)}(t_>, N) 
\label{eq:gen-ising-ff1}\\
\text{low temperature :}\quad \mathcal C^{(-)}(t_<, N; \lambda)&=& \left(1-t_<\right)^{1/4}\left(1+\sum_{p=0}^\infty \lambda^{2p} \,f^{(2p)}(t_<, N) \right)
\label{eq:gen-ising-ff2}
\end{eqnarray}
These generalized form-factor expansions satisfy the same PVI equations as the physical Ising diagonal correlators $\mathcal C^{(\pm)}(t, N)$. Explicitly, define Okamoto PVI sigma functions as follows [do not confuse $\sigma(t, N)$ with the boundary condition parameter $\sigma$ in Jimbo's tau function expansion (\ref{eq:tau-jimbo})]. At  high temperature
\begin{eqnarray}
\sigma^{(+)}(t_>, N)\equiv t_>(t_>-1) \frac{d}{dt_>}\, \ln \, \mathcal C^{(+)}(t_>, N)-\frac{1}{4}
\label{eq:sigma-high}
\end{eqnarray}
and at low temperature define
\begin{eqnarray}
\sigma^{(-)}(t_<, N)\equiv t_<(t_<-1) \frac{d}{dt_<}\, \ln \, \mathcal C^{(-)}(t_<, N)-\frac{t_<}{4}
\label{eq:sigma-low}
\end{eqnarray}
With these definitions, $\sigma^{(\pm)}(t, N)$ both satisfy the Okamoto form of PVI \cite{jimbo-miwa,okamoto}:
\begin{eqnarray}
\left(t(t-1)\sigma^{\prime\prime}\right)^2=N^2\left((t-1)\sigma^\prime-\sigma\right)^2-4\sigma^\prime\left((t-1)\sigma^\prime-\sigma-\frac{1}{4}\right)\left(t\, \sigma^\prime-\sigma\right)
\label{eq:okamoto-p6}
\end{eqnarray}
Note that this Okamoto PVI equation (\ref{eq:okamoto-p6}) is invariant under the duality transformation: $t\to s=\frac{1}{t}$, $\sigma(t)\to \Sigma(s)=\frac{1}{t}\sigma(t)$.

From the perspective of resurgence, we recognize the generalized form-factor expansions in (\ref{eq:gen-ising-ff1}, \ref{eq:gen-ising-ff2}) as {\it trans-series expansions} of the Ising correlators,
with $\lambda$ being the trans-series parameter for the PVI tau function. Solving the Okamoto PVI equation (\ref{eq:okamoto-p6}) with a trans-series ansatz for $\sigma(t, N; \lambda)$, with trans-series parameter $\lambda$, corresponds to the generalized form-factor expansions  (\ref{eq:gen-ising-ff1}, \ref{eq:gen-ising-ff2}) for the correlator $\mathcal C(t, N; \lambda)$, via the relations (\ref{eq:sigma-high}, \ref{eq:sigma-low}). As usual, the trans-series parameter is associated with a choice of boundary conditions, and it also counts instanton sectors. The form-factor functions $f^{(2p+1)}(t_>, N)$ and $f^{(2p)}(t_<, N)$, for instanton sectors labeled by $p$, are therefore related by resurgent relations for different sector labels $p$. The particular boundary conditions for the Ising correlators arise with the trans-series parameter choice $\lambda=1$. This shows again that even though both the high temperature and low temperature diagonal correlator expansions are convergent, the diagonal correlators have a natural trans-series expansion. This is because  the PVI equation is {\it nonlinear}, and therefore naturally develops trans-series expansions, even though the actual fluctuation expansions in each sector are convergent.
Analogous behavior is studied in Section \ref{sec:gww-strong} for  the (convergent) strong-coupling expansion of the partition function of the Gross-Witten-Wadia matrix model, whose corresponding Okamoto sigma function satisfies a Painlev\'e III equation, but nevertheless has a natural trans-series expansion.
 
At each order of the trans-series expansions in (\ref{eq:gen-ising-ff1}, \ref{eq:gen-ising-ff2}), we can expand the functional $t$ dependence in a convergent expansion about $t=0$ (i.e., $T=\infty$). From the large order growth of these expansion coefficients, we deduce from Darboux's theorem (see Appendix 2,  Section \ref{app:darboux}) the analytic function multiplying the logarithmic singularity in the vicinity of the critical temperature, at $t=1$. Thus, for each of the form-factor functions $f^{(2p+1)}(t_>, N)$ and $f^{(2p)}(t_<, N)$, the expansion in the vicinity of $t_c$ is encoded in the expansion about $t=0$. This can be seen very explicitly, for example, in  the resurgent expansions of the low order form-factor functions $f^{(1)}(t, N)$ and $f^{(2)}(t, N)$ from (\ref{eq:ising-f1}, \ref{eq:ising-f2}).
The immediate vicinity of the transition temperature can also be probed by a double-scaling limit, relevant for the scaling limit $N\to \infty$ and $T\to T_c$. This limit reduces the PVI system to a PV equation, and with the particular Ising parameters of this PV system, it can also be expressed in terms of a PIII equation \cite{Wu:1975mw,McCoy:1987qp,deift}. These double-scaling solutions are therefore also resurgent.

\section{Resurgence in the Gross-Witten-Wadia Unitary Matrix Model: Painlev\'e III}
\label{sec:gww}

While the Painlev\'e VI equation `contains' all the lower Painlev\'e equations in the sense that they can all be reached from PVI by a coalescence cascade of limits \cite{nist-painleve}, interesting new physical effects  occur due to the merging of singularities. Singularities can change their character, which affects the form of the expansion about that point. In this Section we illustrate this phenomenon with the example of the Gross-Witten-Wadia (GWW) unitary matrix model \cite{gw,wadia,Rossi:1996hs}, which is characterized by the Painlev\'e III equation \cite{Hisakado:1996di,Ahmed:2017lhl} (and in the double-scaling limit  by the Painlev\'e II equation \cite{gw,wadia,marino-matrix}). Like PVI, the PIII equation has a regular fixed singularity at $t=0$, but after the merging of the other two singularities, PIII has an {\it irregular} fixed singularity at $t=\infty$. This has two immediate consequences: the PIII expansion about $t=0$ now has infinite radius of convergence, but the PIII expansion about $t=\infty$ becomes divergent. In the GWW model, the expansion about $t=0$ is the (convergent) strong-coupling expansion, and the expansion about $t=\infty$ is the (divergent) weak-coupling expansion. As expected from the discussion in Section \ref{sec:p6-pnp} connecting PIII and the Mathieu spectral problem, this expansion behavior for the GWW model \cite{Ahmed:2017lhl} is closely analogous to the behavior of the Mathieu equation spectral problem \cite{Basar:2015xna}. 

The GWW matrix model describes $N\times N$ unitary matrices, with the following  partition function, involving a coupling strength $g^2$:
\begin{eqnarray}
Z(g^2, N)=\int_{U(N)} DU \, \exp\left[\frac{1}{g^2} {\rm tr}
\left(U+U^\dagger\right)\right]
\label{eq:z}
\end{eqnarray}
The GWW model has an interesting third order phase transition in the double-scaling limit: $N\to\infty$, with $g^2\to 0$ such that $N g^2$ is fixed \cite{gw,wadia}.
The matrix integral in (\ref{eq:z}) can be evaluated as a Toeplitz determinant, for any $N$ and any coupling $g^2$ (see the review \cite{Rossi:1996hs})
\begin{eqnarray}
Z(t, N) =\det \left[ I_{j-k} \left(\sqrt{t}\right)\right]_{j, k =1, \dots , N}
\label{eq:gww-z-det}
\end{eqnarray}
The entries of the Toeplitz determinant are  modified Bessel functions, $I_j$, evaluated at the inverse coupling $\sqrt{t}$, where 
\begin{eqnarray}
t\equiv \left(\frac{2}{g^2}\right)^2
\label{eq:gww-t}
\end{eqnarray}
The GWW  third order phase transition occurs at the critical value $\frac{N}{\sqrt{t}}=1$.

\subsection{GWW Trans-series at Strong Coupling}
\label{sec:gww-strong}

Strong coupling means small $t$, so for a given $N$ it is straightforward to expand the Bessel functions in the Toeplitz determinant representation (\ref{eq:gww-z-det}) to obtain an expansion with infinite radius of convergence (for a given fixed $N$):
\begin{eqnarray}
Z(t, N) \sim e^{t/4} \left[1 -  \left(\frac{(\sqrt{t}/2)^{N+1}}{(N+1)!}\right)^2  \left(1-\frac{1}{2}\frac{(N+1)\, t}{(N+2)^2}
+\frac{1}{16}\frac{ (2 N+3)\, t^2}{(N+3)^2 (N+2)} +\dots \right)+\dots \right]
\nonumber\\
\label{eq:gww-z-strong}
\end{eqnarray}
Nevertheless, despite this being a convergent expansion, it has a trans-series structure, as  explained below. This convergent trans-series structure is very similar in form to that of the energy gaps in the Mathieu equation, where all powers of  a factor like $\left(\frac{(\sqrt{t}/2)^{N+1}}{(N+1)!}\right)^2$ appear, and these are associated with {\it complex} instantons, from over-the-barrier tunneling in the periodic cosine potential  \cite{dykhne,Basar:2015xna}. These non-perturbative effects in the GWW strong coupling regime are due to eigenvalue tunneling \cite{marino-matrix}, and can be identified with complex saddle configurations in the partition function integral \cite{Buividovich:2015oju,Alvarez:2016rmo}. The similarities with the Mathieu system can be traced to a common underlying PIII structure.

For any $N$, the GWW partition function $Z(t, N)$ is  a tau function for a Painlev\'e III equation. Defining the Okamoto sigma function as
\begin{eqnarray}
\sigma(t, N) \equiv -t\frac{\partial}{\partial t} \ln \, Z(t, N) +\frac{t}{4}
\label{eq:gww-sigma-z}
\end{eqnarray}
this function $\sigma(t, N)$ satisfies the Okamoto PIII$_1^\prime$ equation \cite{okamoto}:
\begin{eqnarray}
\left(t\, \sigma^{\prime\prime}\right)^2+\sigma^\prime (\sigma-t\, \sigma^\prime)(4\sigma^\prime -1) -N^2 \left(\sigma^\prime\right)^2=0
\label{eq:pIII}
\end{eqnarray}
From this equation it is straightforward to obtain the small $t$ expansion
\begin{eqnarray}
	\frac{\sigma(t, N)}{(N+1)} &\sim& \xi\, C_N \, t^{N+1} \left(1 - \frac{1}{2(N+2)}t + \frac{(2N+3) N!}{16 (N+3)!} t^2 -\frac{(2N+5) N!}{96 (N+4)!} t^3 	+\ldots \right) 
	 \nonumber\\
	&&+\xi^2\, C_N^2\, t^{2N+2} \left(1 - \frac{(2N+3)}{2(N+2)^2} t + \frac{(41 + 59 N + 27 N^2 +  4 N^3)}{8 (2 + N)^3 (3 + N)^2}  t^2+ \ldots \right) 
  \nonumber\\
 &&+ \xi^3\, C_N^3 \, t^{3N+3} \left(1 -\frac{(3N+4)}{2(N+2)^2} t
 +\frac{(5 + 3 N) (59 + 89 N + 41 N^2 + 6 N^3)}{16(N+3)^2(N+2)^4}t^2
 + \ldots \right)
 \nonumber\\
 &&  + \ldots
\label{eq:gww-sigma-strong}
\end{eqnarray}
where $\xi$ is the trans-series parameter, and the numerical coefficient $C_N$ is:
\begin{equation} 
	C_N = \left(\frac{1}{2^{N+1} (N+1)!}\right)^2
	\label{eq:ccn}
\end{equation}
We can write the trans-series solution of this Okamoto PIII$_1^\prime$ equation as:
\begin{eqnarray}
\sigma(t, N) =(N+1)\sum_{k=1}^\infty \xi^k 
\left(\frac{1}{2^{N+1}\, (N+1)!}\right)^{2k}
\, t^{k(N+1)}\, \sigma_{(k)}(t, N)
\label{eq:gww-sigma-trans}
\end{eqnarray}
Here $\xi$ is the trans-series parameter. Matching to the GWW model boundary conditions fixes $\xi=1$. The leading $k=1$ term in (\ref{eq:gww-sigma-trans}), which is the first line in (\ref{eq:gww-sigma-strong}), corresponds when converted via (\ref{eq:gww-sigma-z}) to the partition function $Z(t, N)$ to the expansion term displayed in 
(\ref{eq:gww-z-strong}). This is a convergent expansion. However, the {\it full} trans-series in (\ref{eq:gww-sigma-trans}), including all powers of the instanton-counting parameter $\xi$, is a sum over the contributions associated with complex saddles \cite{Buividovich:2015oju,Ahmed:2017lhl,Alvarez:2016rmo}. The fluctuations about the $k^{\rm th}$ saddle sector, $\sigma_{(k)}(t, N)$, are all {\it convergent}. For example,  the one-instanton ($k=1$) fluctuation factor [see the first line in Eq (\ref{eq:gww-sigma-strong})] can be summed to all orders \cite{Ahmed:2017lhl}:
\begin{eqnarray}
\sigma_{(1)}(t, N)=\frac{t}{4}\left(\left(J_N(\sqrt{t})\right)^2-J_{N+1}(\sqrt{t}) J_{N-1}(\sqrt{t})\right)
\label{eq:gww-sigma-strong-1}
\end{eqnarray}
which clearly has infinite radius of convergence for any $N$. 

In terms of the partition function $Z(t, N)$, the trans-series for $\sigma(t, N)$ becomes (with the GWW trans-series parameter choice, $\xi=1$)
\begin{eqnarray}
e^{-t/4} \, Z(t, N)&\sim& 1-
\frac{\left(\frac{t}{4}\right)^{N+1}}{((N+1)!)^2}\left(1-\frac{1}{2}\frac{(N+1)\, t}{(N+2)^2}
+\frac{1}{16}\frac{ (2 N+3)\, t^2}{(N+3)^2 (N+2)} +\dots \right)
\nonumber\\
&&+
\frac{\left(\frac{t}{4}\right)^{2(N+2)}}{((N+2)! (N+3)!)^2}
\left(1 - \frac{(N+2) t}{(N+4)^2} +\dots\right)
\nonumber\\
&&-
\frac{4\left(\frac{t}{4}\right)^{3(N+3)}}{((N+3)! (N+4)! (N+5)!)^2}\left(1 - O(t) +\dots\right)
+\dots
\label{eq:gww-z-strong-full}
\end{eqnarray}
The first line corresponds to the naive small $t$ expansion shown in (\ref{eq:gww-z-strong}), obtained for example by expanding the Toeplitz determinant for various values of $N$, and fitting.
This fluctuation factor can be re-summed as in (\ref{eq:gww-sigma-strong-1}), but this only agrees with the small $t$ expansion of the exact expression to order $O(t^{2N+1})$. The remaining "higher instanton" terms in (\ref{eq:gww-z-strong-full})  give the full expression to all orders, for any $N$. This is a clear example of the non-perturbative trans-series completion of a convergent expansion.
Notice that when $N$ is an integer, as it is in the GWW model, it is difficult to disentangle the trans-series structure in this convergent small $t$ expansion, but in the generic case this is a series in powers of $t$ and powers of $t^N$, which can be distinguished for generic  non-integer $N$. This is analogous to the structure in Jimbo's PVI expansion (\ref{eq:tau-jimbo}).

Indeed, the expansion of $\sigma(t, N)$ coming from the Okamoto PIII$_1^\prime$ equation (\ref{eq:pIII}) generates a strong coupling (small $t$) expansion for the partition function that matches the all-orders $c=1$ conformal  block tau function expansion in (\ref{eq:tau-all-orders}), suitably reduced to PIII. This coalescence reduction has been performed: see Eqs (5.14)-(5.15) of \cite{Gamayun:2013auu} (with some redefinition of constants, and the identification $r\to N$):
\begin{eqnarray}
e^{-t/4} Z(t, N) = \sum_{k=0}^\infty \left(\frac{G(k+1)G(k+N+1)}{G(2k+N+1)}\right)^2 
(-1)^k \left(\frac{t}{4}\right)^{k(N+k)}
\hskip-.5cm \sum_{\lambda, \mu \in \mathbb Y| \lambda_1, \mu^\prime_1\leq k} \mathcal B_{\lambda, \mu} (k, N) \, t^{|\lambda|+|\mu|}
\nonumber\\
\label{eq:gww-p6}
\end{eqnarray}
where the conformal block coefficients are
\begin{eqnarray}
\mathcal B_{\lambda, \mu} (k, N)&=&
\prod_{(i, j)\in \lambda}\frac{(i-j+k)(i-j+k+N)}{h_\lambda^2(i, j) (\lambda^\prime_j+\mu_i-i-j+1+2k+N)^2} \nonumber\\
&&\times 
\prod_{(i, j)\in \mu}\frac{(i-j-k)(i-j-k-N)}{h_\mu^2(i, j) (\lambda_i+\mu_j^\prime -i-j+1-2k-N)^2}
\label{eq:gww-p6-b}
\end{eqnarray}
It is an instructive exercise to confirm that this conformal block sum reproduces the trans-series form in (\ref{eq:gww-z-strong-full}), which is derived directly from the Okamoto PIII$_1^\prime$ equation (\ref{eq:pIII}). Note the characteristic {\it quadratic} dependence on the instanton number $k$ of the power appearing in the instanton factors $\left(\frac{t}{4}\right)^{k(N+k)}$ for the tau function in (\ref{eq:gww-p6}), compared to the {\it linear} dependence on $k$ of the powers appearing in the corresponding expansions (\ref{eq:gww-sigma-strong}, \ref{eq:gww-sigma-trans}) of the sigma function $\sigma(t, N)$. This quadratic dependence for the tau function can be traced back to Jimbo's expansion (\ref{eq:tau-jimbo}) and its all orders form in terms of conformal blocks (\ref{eq:tau-all-orders}), noting the quadratic dependence on the boundary condition parameter $\tilde{\sigma}$ (\ref{eq:tau-jimbo}, \ref{eq:block}), and the resurgent shift $\tilde\sigma\to \tilde\sigma+n$ in (\ref{eq:tau-all-orders}). 

The normalization factors in (\ref{eq:gww-p6}) generate the prefactors of the expansion in (\ref{eq:gww-z-strong-full}):
\begin{eqnarray}
&&\left(\frac{G(k+1)G(k+N+1)}{G(2k+N+1)}\right)^2 
(-1)^k \left(\frac{t}{4}\right)^{k(N+k)}
\nonumber\\
&&\longrightarrow \left\{1,
\frac{-\left(\frac{t}{4}\right)^{N+1}}{((N+1)!)^2},
\frac{+\left(\frac{t}{4}\right)^{2(N+2)}}{((N+2)! (N+3)!)^2},
\frac{-4\left(\frac{t}{4}\right)^{3(N+3)}}{((N+3)! (N+4)! (N+5)!)^2}, \dots \right\}
\label{eq:gww-check}
\end{eqnarray}
where we have used the Barnes G function property
\begin{eqnarray}
\frac{G(x+1)}{G(x)}=\Gamma(x)
\label{eq:barnes}
\end{eqnarray}
together with $G(1)=G(2)=G(3)=1$, $G(4)=2$.

In fact,  the GWW partition function can also be expressed as a sum over partitions in an even simpler form, due to Gessel and Borodin (see \cite{borodin}):
\begin{eqnarray}
Z(t, N)=\sum_{\lambda \in \mathbb Y; \lambda_1\leq N}\left(\frac{{\rm dim}\,\lambda}{|\lambda |!}\right)^2 \left(\frac{t}{4}\right)^{|\lambda|} 
\label{eq:gww-borodin}
\end{eqnarray}
This expression (\ref{eq:gww-borodin}) is an all-orders expansion of the Toeplitz determinant expression in (\ref{eq:gww-z-det}).
For example, when $N=1$ each Young diagram  has a single row, so we obtain the expansion
\begin{eqnarray}
Z(t, 1)=\sum_{n=0}^\infty \frac{1}{(n!)^2} \left(\frac{t}{4}\right)^{n} =I_0(\sqrt{t})
\label{eq:gww-b1}
\end{eqnarray}
When $N=2$ we sum over Young diagrams with one row, and those with two rows, leading to the expansion
\begin{eqnarray}
Z(t, 2)=\sum_{n=0}^\infty \sum_{j=0}^n \frac{(2j-n+1)}{j(j+1)! ((n-j+1)!)^2} \left(\frac{t}{4}\right)^{n} =I_0^2(\sqrt{t})-I_1^2(\sqrt{t})
\label{eq:gww-b2}
\end{eqnarray}
in agreement with (\ref{eq:gww-z-det}) with $N=2$.

\subsection{GWW Trans-series at Weak Coupling}
\label{sec:gww-weak}

Weak coupling means large $t$, so at fixed $N$ we can use the large $t$ asymptotic expansion of the Bessel functions in the Toeplitz determinant representation (\ref{eq:gww-z-det}). At fixed index $j$, the large $x$ resurgent asymptotic expansion of the modified Bessel function involves two exponential terms:
\begin{eqnarray}
I_j(x) \sim \frac{e^{x}}{\sqrt{2 \pi x}} \sum_{n=0}^\infty(-1)^n \frac{\alpha_n(j)}{x^n} \pm i e^{ij\pi} \frac{e^{-x}}{\sqrt{2 \pi x}} \sum_{n=0}^\infty \frac{\alpha_n(j)}{x^n}, \qquad \left|{\rm arg}(x) - \frac{\pi}{2}\right| <\pi
\label{eq:bessel}
\end{eqnarray}
where the fluctuation coefficients are
\begin{eqnarray}
\alpha_n(j) = \frac{1}{8^n n!}\prod_{l=1}^n \big(4 j^2 - (2l-1)^2\big)
=\frac{1}{8^n n!} \frac{(2j+2n-1)!!}{(2j-2n-1)!!}
\end{eqnarray}
Expanding the Toeplitz determinant therefore leads to a trans-series instanton-sum structure \cite{Ahmed:2017lhl}:
\begin{eqnarray}
Z(t, N)\sim Z_0(t, N) \sum_{k=0}^N Z^{(k)}(t, N) e^{-2\, k\, \sqrt{t}} \sum_{n=0}^\infty  \frac{a_n^{(k)}(N)}{t^{n/2}}
\label{eq:z-transseries}
\end{eqnarray}
where 
\begin{eqnarray}
Z_0(t, N)=\frac{G(N+1)}{(2\pi)^{N/2}} e^{N\, \sqrt{t}} t^{-N^2/4}
\label{eq:z0}
\end{eqnarray}
and exponential prefactors $Z^{(k)}(t, N)$ \cite{Ahmed:2017lhl}.
The first few fluctuation expansions are
\begin{eqnarray}
 \sum_{n=0}^\infty \frac{a_n^{(0)}(N)}{t^{n/2}}&=&1+\frac{N}{8}\frac{1}{\sqrt{t}} + \frac{9N^2}{128} \frac{1}{t} + \frac{3N(17N^2+8)}{1024} \frac{1}{t^{3/2}} + \ldots 
\label{eq:zweaka}\\
\sum_{n=0}^\infty \frac{a_n^{(1)}(N)}{t^{n/2}}&=&1-\frac{(N-2)(2N-3)}{8}\frac{1}{\sqrt{t}} + \frac{(4 N^4-36 N^3+129 N^2-220 N+132)}{128}  \frac{1}{t} +  \ldots 
\label{eq:zweakb}
\\
\sum_{n=0}^\infty \frac{a_n^{(2)}(N)}{t^{n/2}}&=&1-\frac{(N-4)(4N-9)}{8}\frac{1}{\sqrt{t}} + 
\frac{(16 N^4-216 N^3+1113 N^2-2552 N+2160)}{128}\frac{1}{t} +\dots
\label{eq:zweakc}
\end{eqnarray}
 The resurgent properties of the weak-coupling trans-series expansion (\ref{eq:z-transseries}) have been studied in \cite{Ahmed:2017lhl}. The large-order growth of the fluctuation coefficients in a given instanton sector are quantitatively related to the low-order coefficients of the fluctuations in neighboring instanton sectors. For example, at large perturbative fluctuation order, the coefficients in the $k=0$ sector grow as the expansion order $n\to\infty$ as
\begin{eqnarray} 
a^{(0)}_n(N) &\sim& \frac{2^{N}}{\pi (N-1)!}  \frac{ (n+N-3)!}{2^n}
 \left[1-\frac{(N-2)(2N-3)}{8}\frac{2}{(n+N-3)}\right. \\
&&\left. + \frac{(4 N^4-36 N^3+129 N^2-220 N+132)}{128} \frac{2^2}{(n+N-3)(n+N-4)} + \dots \right]
 \nonumber
\label{eq:gww-weak-a0}
\end{eqnarray}
In this expression for the large order behavior of the coefficients of the zero-instanton series (\ref{eq:zweaka}) we recognize the low order coefficients of the one-instanton series (\ref{eq:zweakb}). These large-order/low-order relations hold for any $N$. Similarly, the large-order growth of the fluctuation coefficients in the one-instanton sector is given by  \cite{Ahmed:2017lhl}:
\begin{eqnarray}
a^{(1)}_n(N) &\sim& -\frac{2^{N-1}}{\pi (N-2)!}  \frac{ (n+N-5)!}{2^n} \left[1-\frac{(N-4)(4N-9)}{8}\frac{2}{(n+N-5)}  \right.  \\
&& \left.+ \frac{(16 N^4-216 N^3+1113 N^2-2552 N+2160)}{128}\frac{2^2}{(n+N-5)(n+N-6)}
+\dots\right]
\nonumber
	\label{eq:gwww-weak-a1}
\end{eqnarray}
in which we recognize the low-order terms in the $k=2$ sector in (\ref{eq:zweakc}). These generic resurgent large-order/low-order relations hold for any fixed $N$.

This weak-coupling expansion can be identified with the large $t$ expansion in equations (A.30)-(A.31) of the PIII$_1^\prime$ tau function
in \cite{Bonelli:2016qwg}, where it  is expressed as (with the identification $\theta_*=N/2$, $\theta_\star=-N/2$, $s\to 2 i \sqrt{t}$, and trans-series parameter $e^{i\rho}$):
\begin{eqnarray}
\tau
(t, N)&=&t^{\frac{N^2}{4}}\sum_{k\in \mathbb Z} e^{i\, k\, \rho}\mathcal G(\nu+k, t) 
\label{eq:gww-weak-block1}\\
\mathcal G(\nu, t)&=& C(\nu, t)\left[1+\sum_{n=1}^\infty \frac{D_n(\nu)}{2^n\, t^{n/2}}\right]
 \label{eq:gww-weak-block2}
 \end{eqnarray}
 with normalization factors $C(\nu, t)$ \cite{Bonelli:2016qwg}.
 The first few coefficients of the fluctuation terms in (\ref{eq:gww-weak-block1}, \ref{eq:gww-weak-block2}) can be expressed as:
 \begin{eqnarray}
 D_1(\nu)&=& -\nu ^3+\frac{1}{4} \nu  \left(N^2-2\right) \\
 D_2(\nu)&=& \frac{\nu ^6}{2}+\frac{1}{4} \nu ^4 \left(7- N^2\right)+\frac{1}{32} \nu ^2
   \left(N^2-10\right) \left(N^2-6\right)+\frac{1}{64} N^2 \left(N^2-12\right)
 \end{eqnarray}
With the further  identification $\nu=-N/2$ these reduce to
  \begin{eqnarray}
 D_1\left(-\frac{N}{2}\right)&=&\frac{N}{4}\\ 
 D_2\left(-\frac{N}{2}\right)&=& \frac{9 N^2}{32}
 \end{eqnarray}
 which match the first two terms in the zero-instanton expansion GWW expansion (\ref{eq:zweaka}) above.
 
 But more interesting from the resurgence perspective is that the fluctuation coefficients in the  higher instanton sectors can be obtained from those in the $k=0$ sector, simply by a shift of $\nu$. Note for example, that shifting $\nu\to \nu+1$, and then identifying $\nu=-\frac{N}{2}$, yields
 \begin{eqnarray}
 D_1\left(-\frac{N}{2}+1\right)&=& -\frac{(N-2)(2N-3)}{4}\\
 D_2\left(-\frac{N}{2}+1\right)&=& \frac{(4 N^4-36 N^3+129 N^2-220 N+132)}{32} 
 \end{eqnarray}
 which match the first two terms in the one-instanton expansion GWW expansion (\ref{eq:zweakb}) above.
 Similarly, shifting $\nu\to\nu+2$, and then identifying $\nu=-\frac{N}{2}$, yields
 \begin{eqnarray}
 D_1\left(-\frac{N}{2}+2\right)&=& -\frac{(N-4)(4N-9)}{4}\\
 D_2\left(-\frac{N}{2}+2\right)&=& \frac{(16 N^4-216 N^3+1113 N^2-2552 N+2160)}{32} 
 \end{eqnarray}
 which match the first two terms in the two-instanton expansion GWW expansion (\ref{eq:zweakc}) above. 

Thus, the conformal block expansion (\ref{eq:gww-weak-block1}, \ref{eq:gww-weak-block2})  gives the all-orders trans-series expansion in a concise combinatorial form, and demonstrates the resurgent structure via the simple shift of the boundary condition parameter $\nu=-\frac{N}{2}$ by integer shifts according to the instanton number.

\section{Resurgence in Full Counting Statistics of 1 Dimensional Fermions: Painlev\'e V}
\label{sec:fcs}

It is well known that the Painlev\'e equations play a key role in the physics of one-dimensional fermionic and bosonic systems \cite{jimbo-bose,keating,forrester-book}. Here we briefly mention one particular such application, and discuss the physical significance of the resurgent structure of the associated asymptotic expansion.

Consider the full counting statistics \cite{levitov} for free one dimensional fermions. This quantity is the generating function of the cumulants:
\begin{eqnarray}
\chi(x, \kappa)\equiv \langle e^{2\pi \, i\, \kappa\, \hat{Q}}\rangle
\label{eq:fcs}
\end{eqnarray}
Here $\hat{Q}$ is the charge operator, quadratic in the fermion creation and annihilation operators. The parameter $x$ is related to the Fermi energy, by a suitable scaling, and $\kappa$ is the generating function parameter. For free one dimensional fermions, this function $\chi(x, \kappa)$ is a tau function for Painlev\'e V. Explicitly,  if we define the corresponding Okamoto sigma function:
\begin{eqnarray}
\sigma(x, \kappa)\equiv x\frac{\partial}{\partial x} \ln\, \chi(x, \kappa)
\label{eq:sigma5}
\end{eqnarray}
then $\sigma(x, \kappa)$ satisfies an Okamoto Painlev\'e V equation (see \cite{Ivanov:2011ab} and references therein):
\begin{eqnarray}
\left(x\, \sigma^{\prime\prime}\right)^2+4(x\, \sigma^\prime -\sigma)\left(x\, \sigma^\prime-\sigma+\left(\sigma^\prime\right)^2\right)=0
\label{eq:sigma-p5}
\end{eqnarray}
In the large $x$ limit  the following asymptotic expansion was conjectured and verified to high order \cite{Ivanov:2011ab}:
\begin{eqnarray}
\chi(x, \kappa)&=& \sum_{n=-\infty}^\infty \chi_*(x, \kappa+n) 
\label{eq:chi-exp1}\\
{\rm where}\quad \chi_*(x, \kappa)&=& \left(G(1+\kappa)G(1-\kappa)\right)^2\frac{e^{2i\kappa \, x}}{(2x)^{2\kappa^2}}\sum_{k=0}^\infty \frac{p_k(\kappa)}{(i\, x)^k}
\label{eq:chi-exp2}
\end{eqnarray}
The expansion coefficients $p_k(\kappa)$ are polynomials in $\kappa$. These polynomials can be generated recursively in an algorithmic manner using this ansatz in (\ref{eq:sigma5}) and (\ref{eq:sigma-p5}).

We recognize the expansion (\ref{eq:chi-exp1}) as a trans-series expansion for the tau function $\chi(x, \kappa)$ of the Painlev\'e V system. The "periodicity" in $\kappa$ of the expansion (\ref{eq:chi-exp1}) is manifest by the sum over all integer shifts $\kappa\to\kappa+n$, and the consistency of this ansatz was verified to high order in \cite{Ivanov:2011ab}. Moreover, it was pointed out in \cite{Ivanov:2011ab} that  periodicity in $\kappa$ is physically {\it required} due to the integer nature of the charge $\hat{Q}$ in the definition of the full counting statistics (\ref{eq:fcs}).

As in the case of Painlev\'e III for the Gross-Witten-Wadia model discussed in Section \ref{sec:gww}, the reduction from Painlev\'e VI involves the merging of two regular singular points, resulting in an irregular singular point at infinity. Therefore, the large $x$ expansion  of the Painlev\'e V tau function in (\ref{eq:chi-exp1}) is divergent. For PV there is a divergent conformal block expansion at $x\sim\infty$, and a convergent conformal block expansion at $x\sim 0$. The divergent $c=1$ conformal block expansion for PV has been studied in \cite{Bonelli:2016qwg}, Appendix A, where it is expressed as
\begin{eqnarray}
\tau_V(x)&\sim& \sum_{n=-\infty}^\infty e^{i\, n\, \rho} {\mathcal G}(x, \nu+n) \qquad, \quad x\to \infty
\label{eq:tauV1}\\
{\rm where} \quad {\mathcal G}(x, \nu) &=& C(\nu)\frac{e^{2 i \nu x}}{x^{2\nu^2} }\sum_{k=0}^\infty \frac{D_k(\nu)}{(i\, x)^k}
\label{eq:tauV2}
\end{eqnarray}
where $C(\nu)$ are known normalization factors, and $D_k(\nu)$ are polynomials in $\nu$ which can be generated recursively. 

Comparing (\ref{eq:tauV1}, \ref{eq:tauV2}) with (\ref{eq:chi-exp1}, \ref{eq:chi-exp2}), we see that the boundary conditions for the full-counting statistics solution $\chi(x, \kappa)$ fixes the trans-series parameter (instanton counting parameter) $e^{i\rho}$, and the other trans-series parameter $\nu$ plays the role of the generating function parameter $\kappa$. One can further check that the fluctuation polynomials $p_k(\kappa)$ and $D_k(\nu)$ match. Therefore, the resurgent trans-series structure of the $c=1$ conformal block expansion guarantees the physical condition of periodicity under integer shifts of $\kappa$. Physically, the resurgent sum over instanton and anti-instanton sectors in (\ref{eq:tauV1}, \ref{eq:tauV2}) is identified with the sum over different Fisher-Hartwig branches in (\ref{eq:chi-exp1}, \ref{eq:chi-exp2}).

On the other hand, the expansion at $x=0$ is convergent. Nevertheless, it also has a {\it convergent} trans-series conformal block expansion. This is inherited from the convergent Painlev\'e VI all-orders $c=1$ conformal block expansion in (\ref{eq:tau-all-orders}), suitably reduced from Painlev\'e VI to Painlev\'e V. This scaling reduction from PVI to PV has been carried out in \cite{Gamayun:2013auu} Section 4.2. The structural form of the tau function at small $x$ is
\begin{eqnarray}
\tau_V(x)&\sim&  \sum_{n=-\infty}^\infty s_V^n\, x^{(\nu+n)^2}\, {\mathcal C}(\nu+n)\, {\mathcal B}(x, \nu+n)\qquad , \quad x\to 0
\label{eq:tauV10}\\
{\rm where} \quad {\mathcal B}(x, \nu) &=& \sum_{\lambda, \mu \in \mathbb Y} {\mathcal B}_{\lambda, \mu}(\nu) x^{|\lambda|+|\mu|}
\label{eq:tauV20}
\end{eqnarray}
Here ${\mathcal C}(\nu)$ is a normalization function, whose explicit form is given in \cite{Gamayun:2013auu}, and the coefficients 
${\mathcal B}_{\lambda, \mu}(\nu)$ in the conformal block expansion are expressed in terms of pairs of Young diagrams $\lambda$ and $\mu$. For this particular physical problem, the boundary condition is such that the trans-series parameter $s_V$ takes the value 
\begin{eqnarray} 
s_V=\left(\frac{e^{2 i\kappa}-1}{\pi}\right)
\label{eq:trans-pv}
\end{eqnarray}
and the other boundary condition parameter, written as $\nu$ in (\ref{eq:tauV10}), vanishes on the physical solution. In this limit, the physical periodicity under integer shifts of $\kappa$ is manifest in the fact that the trans-series parameter is a periodic function of $\kappa$.

These PV tau function expansions at large and small $x$ correspond to the those of the full-counting statistics function $\chi(x, \kappa)$. For the associated Okamoto sigma function defined in (\ref{eq:sigma5}), the small $x$ boundary condition is \cite{Ivanov:2011ab}
\begin{eqnarray}
\sigma(x, \kappa) \sim \left(\frac{e^{2 i\kappa}-1}{\pi}\right)x-\left(\frac{e^{2 i\kappa}-1}{\pi}\right)^2 x^2+O(x^3) 
\label{eq:s50}
\end{eqnarray}
And one can verify that the Okamoto PV equation (\ref{eq:sigma-p5}) is solved by a convergent small $x$ expansion
\begin{eqnarray}
\sigma(x, \kappa)\sim \sum_{n=1}^\infty \left(\frac{e^{2 i\kappa}-1}{\pi}\right)^n\, x^n F_n(x^2) \qquad, \quad x\to 0
\label{eq:s50-expansion}
\end{eqnarray}
where the convergent fluctuation functions are $F_n(x^2)=\sum_{k=0}^\infty f_{n, k} \, x^{2k}$, which can be generated recursively. This small $x$ expansion can alternatively be expressed as a sum over $x$ with coefficients that are polynomials in $\left(\frac{e^{2 i\kappa}-1}{\pi}\right)$:
\begin{eqnarray}
\sigma(x, \kappa)\sim \sum_{m=1}^\infty x^m \, P_m\left(\left[\frac{e^{2 i\kappa}-1}{\pi}\right]\right) \qquad, \quad x\to 0
\label{eq:s50-expansion2}
\end{eqnarray}

\section{Conclusions}

This paper emphasizes that the resurgent structure of Painlev\'e equations is particularly transparent when expressed in terms of tau functions and their physical conformal block expansions. This leads to three main types of resurgence. The most general relates expansions about different critical points. The second relates different instanton sectors in the expansion about a given critical point. The third relates the fluctuations about different saddle sectors in associated spectral problems. For Painlev\'e VI the expansions about different critical points are all convergent, but there is still a clear trans-series structure. After coalescence of singularities to lower Painlev\'e equations, some expansions become divergent, thereby changing the form of the trans-series.
We have illustrated these phenomena with various physical applications, for the 2d Ising model (Painlev\'e VI), the Gross-Witten-Wadia unitary matrix model (Painlev\'e III), and the scaled full-counting statistics of one dimensional fermions (Painlev\'e V).

\section{Appendix 1:  Weierstrass Forms of Heun and Painlev\'e VI  Equations}
\label{app:p6-heun-appendix}

It is a standard result that the Heun equation can be transformed from its  ``normal form"   to Weierstrass elliptic form  (see \href{https://dlmf.nist.gov/31}{https://dlmf.nist.gov/31}). The standard normal form is
\begin{eqnarray}
\frac{d^2 w}{dz^2}+\left(\frac{c}{z}+\frac{d}{z-1}+\frac{e}{z-x}\right)\frac{dw}{dz}+\frac{a\, b\, z -q}{z(z-1)(z-x)}\, w=0
\label{eq:heun}
\end{eqnarray}
Note that $a+b+1=c+d+e$. Make the following changes of variable from $z$ to $\xi$, and function from $w(z)$ to $W(\xi)$:
\begin{eqnarray}
x&=& \frac{1}{k^2}\equiv \frac{e_1-e_3}{e_2-e_3}\\
z&=&{\rm sn}^2\left(i\, {\mathbb K}^\prime+(e_1-e_3)^{1/2}\, \xi; \frac{e_2-e_3}{e_1-e_3}\right) \\
w&=&\left({\mathcal P}(\xi)-e_3\right)^{(1-2 c)/4} \left({\mathcal P}(\xi)-e_2\right)^{(1-2 d)/4}
 \left({\mathcal P}(\xi)-e_1\right)^{(1-2 e)/4}\, W(\xi)
\label{eq:heun-conversions}
\end{eqnarray}
Then the Heun equation takes the Schr\"odinger-like form:
\begin{eqnarray}
\frac{d^2 W}{d\xi^2}+\left( H+\sum_{k=0}^3 b_k\, {\mathcal P}(\xi+\omega_k)\right) W(\xi)=0
\label{eq:heun-p-form}
\end{eqnarray}
Here, the $b_k$ coefficients are expressed in terms of the parameters $a, b, c, d, e$ of the Heun equation, and $H$ is expressed in terms of the accessory parameter $q$ of the Heun equation, together with  the parameters $a, b, c, d, e$ of the Heun equation (see \href{https://dlmf.nist.gov/31.2.E11}{https://dlmf.nist.gov/31.2.E11}).

It is also possible to convert the PVI equation to a Weierstrass elliptic form by the following transformations \cite{guzzetti,manin,brezhnev}. Change the variable from $t$ to $\tau$, and the function from $y(t)$to $u(\tau)$, as follows:
\begin{eqnarray}
t\equiv \frac{e_3-e_1}{e_2-e_1} \qquad, \qquad y\equiv \frac{1}{e_2-e_1} \left[{\mathcal P}(u; \{1,  \tau\})-e_1\right] 
\label{eq:p6-conversion-2}
\end{eqnarray}
The half-periods of the Weierstrass $\mathcal P$ function are
\begin{eqnarray}
\tau \equiv 2\pi i t\quad, \quad \omega_1=\frac{1}{2} \quad, \quad \omega_2=\frac{1}{2}(1+\tau) \quad, \quad \omega_3=\frac{1}{2}\,\tau \quad, \quad \omega_0=0
\label{eq:p6-parameters}
\end{eqnarray}
Then the PVI equation can be written as
\begin{eqnarray}
\ddot{u} &=& \sum_{k=0}^3 \nu_k\, {\mathcal P}^\prime(u+\omega_k;\{1, \tau\})
\label{eq:p6u}
\end{eqnarray}
where $\cdot$ denotes $\frac{d}{d\tau}$, and we identify the parameters $\nu_k$ in (\ref{eq:p6u}) with the PVI parameters in (\ref{eq:p6})  as
\begin{eqnarray}
\nu_0=\alpha\quad, \quad \nu_1 =-\beta\quad, \quad \nu_2=\gamma\quad, \quad \nu_3=-\delta+\frac{1}{2}
\label{eq:p6-params}
\end{eqnarray}
Note that this form (\ref{eq:p6u}) of the PVI equation is the classical Newtonian equation of motion, $\ddot u=-\frac{\partial V(u, t)}{\partial u}$, with an explicitly time-dependent classical potential:
\begin{eqnarray}
V_{\rm VI}(u, t)&=&-\sum_{k=0}^3 \nu_k\, {\mathcal P}(u(t)+\omega_k;\{1, 2\pi i t\})
\label{eq:p6uv}
\end{eqnarray}
The potential depends on $t$ both through the dependence of $u(t)$ on $t$, and also through the parameter $\tau\equiv 2\pi i t$, which determines the periods of the elliptic P function.

Thus the  potential in the Weierstrass form (\ref{eq:heun-p-form}) of the Heun equation coincides with the potential whose classical equation of motion yields the Weierstrass form (\ref{eq:p6u}) of the PVI equation. This fact is important for the resurgent correspondence, discussed in Section \ref{sec:p6-classical-blocks}, between the PVI and Heun equations.

%

\section{Appendix 2: Darboux and Resurgence Relations}
\label{app:darboux}

Darboux's theorem states that for a convergent series expansion, the large-order growth of the expansion coefficients about a point (say $z=0$) is directly related to the behavior of the expansion in the vicinity of a nearby singularity \cite{fisher,henrici,guttmann}. For example, suppose
\begin{eqnarray}
f(z)\sim  \phi(z)\, \left(1-\frac{z}{z_0}\right)^{-g}+\psi(z) \qquad, \quad z\to z_0 
\label{eq:darboux1}
\end{eqnarray}
where $\phi(z)$ and $\psi(z)$ are analytic near $z_0$. 
Then the Taylor expansion coefficients of $f(z)$ near the origin have large-order growth
\begin{eqnarray}
b_n\sim \frac{1}{z_0^n}\begin{pmatrix}
n+g-1\\ n
\end{pmatrix} \left[ \phi(z_0)- 
\frac{(g-1)\, z_0\, \phi^\prime(z_0)}{(n+g-1)}
+
\frac{(g-1)(g-2)\, z_0^2\, \phi^{\prime\prime}(z_0)}{2! (n+g-1)(n+g-2)}\, -\dots \right]
\label{eq:darboux2}
\end{eqnarray}
Thus,  leading and subleading  large-order behavior terms determine the Taylor expansion of the analytic function $\phi(z)$ which multiplies the branch-cut factor in (\ref{eq:darboux1}).
If the singularity is logarithmic,
\begin{eqnarray}
f(z)\sim  \phi(z)\, \ln\left(1-\frac{z}{z_0}\right)+\psi(z) \qquad, \quad z\to z_0 
\label{eq:darboux3}
\end{eqnarray}
where $\phi(z)$ and $\psi(z)$ are analytic near $z_0$, 
then the Taylor expansion coefficients of $f(z)$ near the origin have large-order growth
\begin{eqnarray}
b_n\sim \frac{1}{z_0^n}\cdot \frac{1}{n} \left[\phi(z_0) - \frac{z_0\, \phi^\prime(z_0)}{(n-1)} +\frac{z_0^2\, \phi^{\prime\prime}(z_0)}{(n-1)(n-2)} -\dots \right]
\label{eq:darboux4}
\end{eqnarray}
Once again, the large-order behavior of the convergent expansion coefficients determines the nature of the singularity, and the fluctuations about it.

The generic Berry-Howls resurgence relations \cite{berry-howls} which connect the large-order growth of the coefficients of a divergent series to the low-order expansion coefficients of the function at its nearby singularities can be understood as an application of Darboux's theorem to the (convergent) Borel transform function (in the Borel variable). By contrast here, since the expansions in the physical variable are convergent, we apply Darboux's theorem directly to the physical function itself, rather than to its Borel transform. It can be viewed as  a demonstration of "resurgence", because the behavior at a certain point can be used to deduce properties of the expansion about the nearest singularity.

\section{Acknowledgements}
This material is based upon work supported by the U.S. Department of Energy, Office of Science, Office of High Energy Physics under Award Number DE-SC0010339.  I thank Anees Ahmed, G\"ok\c ce Ba\c sar, Carl Bender, Chris Coger, Ovidiu Costin, Sasha Gorsky, Sergei Gukov, Sergei Lukyanov, Nikita Nekrasov, Norbert Schuch, Mithat \"Unsal  and Frank Verstraete for helpful discussions and correspondence.  Much of this work was done in Fall 2017 at the KITP at UC Santa Barbara during the research program ``Resurgent Asymptotics in Physics and Mathematics''. Research at KITP is supported by the National Science Foundation under
Grant No. NSF PHY-1125915.

\end{document}